\documentclass[11pt]{article}
\usepackage{amsmath,amsthm,amssymb,natbib}
\setcitestyle{citesep={;}}
\usepackage{mathrsfs}
\usepackage{arydshln}
\usepackage{titlesec}
\usepackage{graphics,graphicx}
\usepackage{epstopdf}
\usepackage{enumerate}
\usepackage{bm,color}
\usepackage[colorlinks=true,linkcolor=blue,urlcolor=blue,citecolor=blue]{hyperref}
\usepackage{float}
\usepackage{subfigure}
\usepackage{amsfonts,latexsym}
\usepackage{multirow}
\usepackage[titletoc,toc,title]{appendix}
\usepackage{autobreak}
    \allowdisplaybreaks
 \usepackage{diagbox,array} 
\usepackage{lscape}
\usepackage{multirow}
\usepackage{longtable}
\usepackage{booktabs}

\begin{document}
\renewcommand{\theequation}{\arabic{section}.\arabic{equation}}
\title{\bf Determining the Number of Common Functional Factors with Twice Cross-Validation}
\author{Hui Jiang $^{\dag}$, Lei Huang $^{\ddag,}$\footnote{Corresponding Author.} and Shengfan Wu $^{\dag}$
\
\\ \vspace{0.3cm}\\
\small{$^{\dag}$School of Mathematics and Statistics,}\\
\small{Huazhong University of Science and Technology, China}\\
\small{$^{\ddag}$School of Mathematics,}\\
\small{Southwest Jiaotong University, China}\\
}
\date{}
\maketitle

\newtheorem{Theorem}{Theorem}
\newtheorem{Example}{\it Example}
\newtheorem{Lemma}{Lemma}
\newtheorem{Note}{Note}
\newtheorem{Proposition}{Proposition}
\newtheorem{Corollary}{Corollary}
\newtheorem{Remark}{Remark}
\numberwithin{equation}{section}

\def\colbl{\textcolor{black}}
\def\colb{\textcolor{blue}}
\def\colr{\textcolor{red}}
\def\colg{\textcolor{green}}
\def\colm{\textcolor{magenta}}

\def\beginn{\begin{eqnarray*}}
\def\endn{\end{eqnarray*}}
\def\beginy{\begin{eqnarray}}
\def\endy{\end{eqnarray}}
\def\n{\nonumber}

\def\I{\mathcal{I}}
\def\Y{\mathcal{Y}}
\def\X{\mathrm{X}}
\def\U{\mathcal{U}}
\def\p{\mathrm{p}}
\def\e{\bm{\epsilon}}
\def\bY{\mathbf{Y}}
\def\E{\mathbf{E}}
\def\F{\mathbf{F}}
\def\B{\mathbf{B}}
\def\G{\mathbf{G}}
\def\V{\mathbf{V}}
\def\D{\mathbf{D}}
\def\bLambda{\mathbf{\Lambda}}
\def\cS{\mathcal{S}}
\def\cP{\mathcal{P}}
\def\1{\mathbf{1}}

\baselineskip1.9em
\begin{abstract}
The semiparametric factor model serves as  a vital tool to describe the dependence patterns in the data. It recognizes that the common features observed in the data are actually explained by functions of specific exogenous variables. 
Unlike traditional factor models, where the focus is on selecting the number of factors, our objective here is to identify the appropriate number of common functions, a crucial parameter in this model. In this paper, we develop a novel data-driven method to determine the number of functional factors using cross validation (CV). Our proposed method employs a two-step CV process that ensures the orthogonality of functional factors, which we refer to as Functional Twice Cross-Validation (FTCV). Extensive simulations demonstrate that FTCV accurately selects the number of common functions and outperforms existing methods in most cases. 
Furthermore, by specifying market volatility as the exogenous force, we provide real data examples that illustrate the interpretability of selected common functions in characterizing the influence on U.S. Treasury Yields and the cross correlations between Dow30 returns.

\end{abstract}

\newpage
\
\baselineskip1.89em
\section{Introduction} \label{sec1}
Over the last few decades, researchers have encountered vast quantities of large and highly correlated datasets in various scientific fields. In financial studies, the returns of financial assets may fluctuate in response to market return, market volatility, or time. In medical studies, a variety of atmospheric pollutants, such as nitrogen dioxide, sulphur dioxide, respirable suspended particulates, and ozone are known to affect public health. Meanwhile, even though the multiple responses display their own patterns of variation with the change of specific driving force, generally they tend to share some common features. Therefore, dimension reduction methods are indispensable for constructing appropriate models to accurately extract these shared features.

\par Recently, factor models
have emerged as effective tools for explaining complex phenomena in finance, economics and other disciplines.
These models allow us to capture the variations in high-dimensional variables using a small number of underlying common factors, thereby avoiding an explosion in the number of unknown parameters.
Well-known attempts in this direction include the arbitrage pricing theory \citep{ross1976arbitrage},
multi-factor models \citep{fama1993common},
inflation forecasting with diffusion indices \citep{stock2002macroeconomic}, and the aggregate implications of macroeconomic behavior \citep{forni1997aggregation}.
Unfortunately, they still have limited capability in characterizing certain common features. It is worth highlighting that the similar patterns observed across multiple responses may not solely be driven by specific exogenous variables themselves, but rather stem from the common functional forms of such variables. For instance, the asymmetric effect of market return on pairwise correlations of stock returns is actually explained by an asymmetric function of market return, but not the market variables \citep{jiang2016asymmetric}. Taking this into consideration, our paper employs the semiparametric factor models to effectively characterize the shared features by utilizing common functional factors.

The fundamental issue of factor models is the preliminary identification of the number of common factors, as this directly impacts the complexity of the model. In the context of semiparametric factor models, which offer more flexibility in terms of model structure, determining the appropriate number of common functions presents a significant challenge. An excessively large number may lead to over-fitting, while an extremely small number may result in substantial information loss and prediction bias. Therefore, it is crucial to thoroughly evaluate the trade-offs and choose an optimal number that enables meaningful practical interpretation.

\par Numerous selection methods have been developed for traditional factor models. One category of approaches for determining  the number of factors is to employ the eigen-structure of data matrix. \cite{ye2003using} introduced a bootstrap selection procedure for order determination. \cite{onatski2010determining} suggested a consistent estimator in the approximated factor models, allowing for substantial correlations among idiosyncratic terms. \cite{luo2016combining} combined the eigenvalues and the bootstrap variability of eigenvectors for order determination. However, these methods require strong assumptions regarding the separability of eigenvalues, potentially limiting their applicability in high-dimensional data. \cite{bura2011dimension} developed sequential tests based on the smallest eigen or singular values of the target matrix to choose the number. Although there exists clear statistical interpretation for the sequential testing, the asymptotic expansions of the proposed estimator are quite complicated \citep{li2007directional}. In addition, it is important to recognize that hypothesis testing-based approaches cannot guarantee consistency at a fixed significance level.

\par Currently, cross validation (CV) and information criteria (IC) have attracted considerable attention for determining the number of factors. CV serves as a model evaluation procedure aiming to identify the best model by minimizing prediction errors. In their comprehensive review, \cite{bro2008cross} explored six CV-based approaches and concluded that most of them are unlikely to ensure statistical consistency. Therefore, they focused on a competitive algorithm named expectation maximization (EM) approach, but it seems to be computationally intensive. Another set of approaches relies on information criteria, which not only ensures consistency but also effectively reduces computational burden. The use of information criteria for determining the number of factors was initially considered by \cite{cragg1997inferring}, but this method tends to underperform even in moderately large dimensions. \cite{stock1998diffusion} modified the Bayesian information criterion (BIC) to select the optimal number of factors, but their rule may not be appropriate beyond the forecasting framework. \cite{forni2000generalized} introduced a multivariate variant of the Akaike information
criterion (AIC), but neither the theoretical nor the empirical properties were provided. \cite{bai2002determining} proposed a panel criterion (PC) that has been proven to be consistent in high dimensional settings.
Unfortunately, PC tends to overestimate the actual number of factors under some circumstances. To overcome this problem, several modifications of the PC have been developed in the literature \citep{hallin2007determining,li2017determining}. However, these IC-based methods are usually not fully data-driven because their penalties are functions of some predetermined tuning parameters.
In a recent development, \cite{zeng2019double} proposed a double cross validation (DCV) to estimate the number of factors. DCV employs cross-validation in two directions, ensuring not only statistical consistency under mild conditions but also achieving computational efficiency.

\par Although most of the existing selection methods may be applicable for dynamic factor models, there is a notable scarcity of methods designed for models in which factors are functions of exogenous variables. 
In this paper, we attempt to bridge this gap within the context of functional factor models.
Following the idea of double cross validation, we develop a novel data driven method capable of effectively selecting the number of common functional factors.

\par The remainder of this article is organized as follows. Section \ref{sec2} discusses the semiparametric factor model relying on exogenous variables and presents the associated model assumptions. Section \ref{sec3} illustrates the modification of two existing methods and introduces our proposed method for estimating the number of common functional factors.
Section \ref{sec5} and \ref{sec6} present the results of simulation studies and empirical examples, offering insights into the finite sample performance of our method. The concluding remarks are given in Section \ref{sec7}.

\section{The Semiparametric Factor model} \label{sec2}
To our knowledge, traditional factor models are effective tools to extract common features by providing parsimonious specifications, but they are very rigorous to be applicable to large dimensional functional data. In light of this, and to accommodate the influence of exogenous variables on changes in response variables, we introduce the following semiparametric factor model in this section.
\subsection{Model Assumption}
Consider the multiple nonparametric regression model
\beginy
Y=G(U)+\epsilon, \label{Model0}
\endy
where $Y \in \mathbb{R}^m$ satisfies $E(Y)=0$, $U \in \mathbb{R}$,  $E(\epsilon)=0$.
Let $Y= (Y_1, \cdots, Y_m)^\top$, $G(U) = (G_1(U), \cdots, G_m(U))^\top$, and
$\epsilon= (\epsilon_1, \cdots, \epsilon_m)^\top$, then model \eqref{Model0} could be presented as
\beginy
\begin{cases}
Y_1=G_1(U)+\epsilon_1,\\
\ \vdots\ \ \ \ \ \  \  \vdots \\
Y_m=G_m(U)+\epsilon_m,
\end{cases}
\endy
Typically, the regression functions $G_1(\cdot), \cdots, G_m(\cdot)$ exhibit similar patterns, which originate from the common functional forms among them. Therefore, following the idea of factor models, we could construct the additive model by introducing some common functions as follows,
\beginy
G_s(U)=b_{s1}^0F_1^0(U)+b_{s2}^0F_2^0(U)+...+b_{s,p_0}^0F_{p_0}^0(U),   \ \ s=1,\cdots,m, \label{ModelF}
\endy
thus
\beginy
Y_s=b_{s1}^0F_1^0(U)+b_{s2}^0F_2^0(U)+...+b_{s,p_0}^0F_{p_0}^0(U)+\epsilon_s,   \ \ s=1,\cdots,m, \label{Model1}
\endy
where $\B^0=(b_1^0, \cdots, b_m^0)^\top$ is the loading matrix with $b_s^0=(b_{s1}^0, \cdots, b_{s,p_0}^0)^\top, s=1,\cdots,m$, $F^0(U)=(F_1^0(U), \cdots, F_{p_0}^0(U))^\top$ is the vector of $p_0$ common factors, and $\epsilon= (\epsilon_1, \cdots, \epsilon_m)^\top$ is the idiosyncratic component. With observations at $\{U_i,i=1,2, \cdots ,n\}$, let $\varepsilon_i=(\epsilon_{i1}, \cdots,\\ \epsilon_{im})^\top$, $\Y_i =(y_{i1}, \cdots,y_{im} )^\top$, $\U=(U_1,\cdots,U_n)^\top$, $\bY=(\Y_1, \cdots, \Y_n)^\top$, $\E=(\varepsilon_1, \cdots, \varepsilon_n)^\top$, and $\F^0=(F^0(U_1), \cdots, F^0(U_n))^\top$. The semiparametric factor model \eqref{Model1} can be expressed in the following matrix form:
\beginy
\bY=\F^0{\B^0}^\top+\E.
\endy
In this article, it is assumed that
\beginy\label{condF}
&&E[F_j^0(U)]=0,\ \  j=1,\cdots,p, \n\\
&&E[F_{j_1}^0(U) F_{j_2}^0(U)] =0,\ \  j_1, j_2=1,\cdots,p, \ \ j_1\neq j_2,\\
&&Var(F_1^0) \geq Var(F_2^0) \geq \cdots \geq Var(F_p^0)  \n
\endy
for identification purpose. Additionally, instead of assuming independent idiosyncratic components, we allow for weak dependencies among them.

Unlike traditional factor models, in which the factors represent either observable or latent variables, our factors are actually some unknown common functions. Assumption \eqref{condF} states that these common functional factors are uncorrelated. This suggests that each factor may capture a specific aspect of information, allowing for explicit interpretations of these common functions. Additionally, the decreasing variances of the functional factors imply that a substantial portion of variability can be explained by the first few common functions, leading to a more parsimonious model. Moreover, as cross-sectional independence among $\epsilon_1, \cdots, \epsilon_m$ is usually unrealistic, we assume weak dependence among the idiosyncratic components throughout this article. Consequently, our discussions primarily revolve around the semiparametric factor model defined in \eqref{Model1}.

\subsection{Estimation of Common Factors and Coefficients} \label{sec2-2}
\par According to the definition of model \eqref{Model1}, both the common functions and corresponding coefficients are actually unknown. In order to obtain estimators that satisfy the conditions, we apply the procedures outlined in \cite{jiang2016asymmetric} for estimating common functional factors and loadings.

Firstly, we can estimate each function $G_s(u)$ separately by
\beginy
\widehat{G}_s(u)=\frac{\sum_{i=1}^{n}y_{is} W_{n,h}(U_i-u)}{\sum_{i=1}^{n}W_{n,h}(U_i-u)},\ \ i=1,\cdots, n, \ \ s=1,\cdots,m,
\endy
where $W_{n,h}(U_i-u) = s_{n,h,2}K_h(U_i-u) - s_{n,h,1}K_h(U_i-u)(U_i-u)$, $K_h(U_i-u) = K\big(\frac{U_i-u}{h}\big)/h$, and $s_{n,h,r}=\sum_{t=1}^n K_h(U_i-u)(U_i-u)^r$ for $r= 0, 1, 2$. Here $K(\cdot)$ is a kernel function, and $h$ is the bandwidth.

\par Denote  the covariance matrix of $G(U)$ by $\bLambda=Cov(G(U))=E(G(U) G(U)^\top)$ , the estimator of
$\bLambda$ can be given by
\beginy
\bLambda_{\widehat{\G}}=\frac{1}{n}\widehat{\G}^\top\hat{\G}
\endy
with $\hat\G=(\hat G^\top(U_1), \cdots, \hat G^\top(U_n))^\top$, and $\hat G(U_i) =  (\hat G_1(U_i), \cdots,\hat G_m(U_i))^\top$.
We assume that the eigenvalues of $\bLambda$ satisfy $\lambda_1> \cdots >\lambda_{p_0}>0$, $\lambda_{p_0+1}= \cdots =\lambda_{m}=0$, and the diagonal matrix $E[F^0(U) {F^0}^\top(U)] =diag(\lambda_1, \cdots,\lambda_{p_0}) =\D^*$ is formed by the largest $p_0$ eigenvalues. Let $V_1, \cdots, V_m$ denote the corresponding orthonormal eigenvectors, and define $\V^*=(V_1, \cdots, V_{p_0})$. Moreover, by
employing the eigenvalue-eigenvector decomposition,
the empirical estimates of the eigenvalues of $\bLambda$ and corresponding orthonormal eigenvectors can be obtained through $\bLambda_{\widehat{\G}}$. With working number of factors $p$, the estimates can be denoted as $\widehat{\V}^{*}=(\widehat{V}_1, \cdots ,\widehat{V}_p)$ and $\widehat\D^* = diag(\widehat\lambda_1, \cdots,\widehat\lambda_p)$.
Finally, according to the derivation presented in \cite{jiang2016asymmetric}, the estimators of common functional factors and loadings can be obtained respectively:
\beginy
\widetilde{\B}^p=\widehat{\V}^{*},
\widetilde{\F}^{p}=\widehat{\G}\widehat{\V}^*.    \label{Model3}
\endy

\section{Estimation of the number of common functions} \label{sec3}
Numerous methods have been developed for determining the number of factors, including the approaches based on the eigen-structure of data matrix, information criteria, or cross validation. In this section, we will briefly introduce several existing representative methods and mainly focus on our proposed functional twice cross validation (FTCV). Meanwhile, we have also made certain adjustments to existing approaches in order to make them applicable to our semiparametric factor model.

\subsection{The Panel Criterion \texorpdfstring{$IC_\p$}{}}
In determining the number of factors in a statistical model, \cite{bai2002determining} proposed the use of panel information criteria. These criteria involve penalty functions based on the dimensions of the observations ($n$) and variables ($m$), leading to consistent estimation of the number of factors.
Notably, the panel information criterion ($IC_\p$) demonstrates a higher frequency of correctly selecting the number of common factors compared to other criteria like AIC and BIC.
Furthermore, \cite{li2013selecting} applied the panel information criterion in the selection of the number of principal components in functional data analysis. Hence, we adopt the concept of $IC_\p$ in our semiparametric factor model, considering it as a representative of IC-based methods for comparison.

In this article, we consider the following class of panel information criteria ($IC_\p$):
\beginy
IC_\p(p)=\log(\widehat{\sigma}_{[p]}^2)+pg_{n,m}
\endy
where
$$\widehat{\sigma}_{[p]}^2=\frac{1}{nm}\sum_{i=1}^n\sum_{s=1}^m \left(y_{is}-\hat{b}_{s1}\widehat{F}_1(U_i)-\hat{b}_{s2}\widehat{F}_2(U_i)-...-\hat{b}_{sp}\widehat{F}_p(U_i)\right)^2,$$
is the error variance estimator and
$$g_{n,m}=(\frac{n+m}{nm})\log{\frac{nm}{n+m}}$$
is the penalty function. Finally, the number of common functional factors is selected by
\beginy
\widehat{p}_{IC_\p}=\mathop{\arg\min}_{p}IC_\p(p).
\endy
Actually, there are other options for the penalty $g_{n,m} $\citep{bai2002determining}, but they tend to perform similarly in practice. Therefore, we only focus on this criterion in this category for further comparison.

\subsection{The \texorpdfstring{$Ladle$}{} Estimator}
Taking into account the eigen-structure of a random matrix, it is evident that when the eigenvalues are widely spaced, the corresponding eigenvectors demonstrate typically small variability. Conversely, when the eigenvalues are closely packed, the corresponding eigenvectors tend to exhibit significant variability.
Based on this discovery, \cite{luo2016combining} proposed a method called $Ladle$ to determine the number of factors with the combination of both. For our problem, we can compute the estimated eigenvalues through $\bLambda_{\widehat{\G}}$. Let $\widehat{D}^*=diag(\hat{\lambda}_1,\hat{\lambda}_2,...,\hat{\lambda}_{p_{max}+1})$, the scaling function of eigenvalues (Rescaled scree plot) can be obtained by
\beginy
{\O}_n(p)=\frac{\hat{\lambda}_{p+1}}{1+\sum_{s=1}^{pmax+1}\hat{\lambda}_s},  (p=0,1,2,...,p_{max}).
\endy
Next, we could generate a set of bootstrap samples ${U_1^*,U_2^*,...,U_n^*}$ , and estimate $\bLambda$ by $\bLambda^*$ based on each bootstrap sample. For each bootstrap sample with factor number $p$, we could take the first $p$ eigenvectors of $\bLambda^*$ to form the matrix $\mathbf{B}_{p,i}^*$ ($i=1,\cdots,n$). We can also obtain $\widehat{\mathbf{B}}^p$ on the full sample using a similar procedure. Then, the variation of $\mathbf{B}_{p,i}^*$ ($i=1,\cdots,n$) around the  estimated $\widehat{\mathbf{B}}^p$ can be measured by
\beginy
f_n^0(p)=\begin{cases}
0,&   p=0,\\
n^{-1}\sum_{i=1}^n\{1-|det(\widehat{\mathbf{B}}^{p^T}\mathbf{B}_{p,i}^*)\}, &
p=1,2,...,p_{max}.
\end{cases}
\endy
Hence, we can obtain the rescaled bootstrap eigenvector variability by
\beginy
f_n(p)=\frac{f_n^0(p)}{1+\sum_{k=0}^{p_max} f_n^0(k)}.
\endy
Finally, by combining the above scale-free quantities, we define
\beginy
g_n(p)={\O}_n(p)+f_n(p),
\endy
thus the estimator of the number of common factors of the adjusted method is obtained by
\beginy
\widehat{p}_{Ladle}=\mathop{\arg\min}_{p}g_n(p).
\endy

Traditional selection methods based on eigen-structure for estimating the number of common factors often rely on either eigenvalues or eigenvectors.  In contrast, $Ladle$ collects information from both eigenvectors and eigenvalues, the combination of which thus provides a scale free estimator and improves the accuracy.
\subsection{The New Method: Functional Twice Cross Validation} \label{sec3-3}
\cite{zeng2019double} introduced a computationally efficient double cross validation (DCV) method, which can guarantee the consistency under mild conditions and avoid predetermining the penalty functions. In the context of our model \eqref{Model1}, the common features are actually explained by specific common functions that adhere to certain conditions. In order to effectively select the number of common functions, we propose a method called functional twice cross validation (FTCV). This approach utilizes the information provided by the exogenous variable $U$ to make informed decisions in the selection process.

Our aim is to determine the number of common functions $p_0$, and our basic idea is to implement the selection based on the prediction error in a twice cross-validated manner. This leads us to develop a multi-stage procedure that takes into account the assumption of uncorrelated unknown common functional factors. To illustrate the procedure, we first rewrite the model elementwise as
\beginy
y_{is}=G_s(U_i) +\epsilon_{is}= {F^0}^\top(U_i)b_s^0+\epsilon_{is},   \label{Model2}
\endy
where both $F^0(\cdot)$ and $b_s^0$ are unobservable. With the working number $p$ of common functions, the details of the multi-stage algorithm are described as follows.

\textbf{\emph{Step 1 }Estimation of coefficients with leave-observation-out CV.}
We first divide the rows of $\bY$ into $K$ ($1<K\leq n$) folds, denoting them as $R_1,...,R_K$ respectively. Let $n_k=\#R_k$ be the number of elements in fold $R_k$ and let $\bY_{-R_k}$, $\U_{-R_k}$ be sub-matrices of $\bY$, $\U$ with rows in $R_k$ removed. We then apply the estimation procedure presented in Section \ref{sec2-2} to data $\bY_{-R_k}$ and $\U_{-R_k}$, This involves computing the kernel estimator $\widehat\G_{-R_k}$, which allows us to obtain the corresponding estimators $\widehat{\F}^{k,p}$ and $\widehat{\B}^{k,p}$ for the common functional factors and coefficients. It is important to note that the matrices $\bY_{-R_k}$, $\widehat\G_{-R_k}$, $\widehat{\F}^{k,p}$ and $\widehat{\B}^{k,p}$ have sizes $(n-n_k)\times m$, $(n-n_k)\times m$, $(n-n_k)\times p$, $m\times p$, respectively. The superscript $k,p$ are used to emphasize that these estimators are obtained from the data with rows in $R_k$ removed, using a working number of factors $p$.

\textbf{\emph{Step 2 }Estimation of orthogonal functional factors with leave-variable-out CV.} In this step, we substitute $b_s^0$ with \eqref{Model2} by $\hat b_s^{k,p}$  and reformulate the model as follows:
\beginy
y_{is}={F^0}^T(U_i)\hat{b}_s^{k,p}+\epsilon_{is}^{'},\ \ i\in R_k, \ \ s=1,\cdots,m,\label{Model4}
\endy
where $\epsilon_{is}^{'}$ represents the regression error resulting from the substitution for $\epsilon_{is}$.
We predict the elements in $\Y_i$, denoted as $y_{is}$ for $s=1,\cdots,m$, thus they share the common $F^0(U_i)$. In this case, all elements in $\Y_i$ can be treated as $m$ observations of an univariate response variable, and $\hat{b}_s^{k,p}$ are known which could be viewed as the regressors. Consequently, the elements in $F^0(U_i)$ can be regarded as the regression coefficients that need estimation. However, different from ordinary regression analysis, orthogonal restrictions need to be imposed on such $F^0(U_i)$ $(i=1,\cdots,n)$ due to the conditions specified in \eqref{condF}. In view of this, we take advantage of the orthogonality between explanatory variables and residuals in least square regression to compute the estimates of $F^0(U_i)$.

By leaving $y_{is}$ out in $R_k$ group, we denote $\bY_{R_k,-s}$ and $\U_{R_k,-s}$ be the sub-matrix of $\mathbf{Y}_{R_k}$ and $\U_{R_k}$ with $s_{th}$ column being removed and let $\hat{\mathbf{B}}^{k,p}_{-s}=(\hat{b}_1^{k,p}, \cdots, \hat b_{s-1}^{k,p},\hat b_{s+1}^{k,p},\cdots, \hat b_m^{k,p})^\top$.
First, we estimate $F_1^0(U_i)$ by regressing the first row of $\bY_{R_k,-s}$ on the vector $(\hat{b}_{11}^{k,p}, \cdots, \hat b_{s-1,1}^{k,p},\\\hat b_{s+1,1}^{k,p},\cdots, \hat b_{m,1}^{k,p})$ with the OLS method, i.e.
\beginn
\hat{F}^{k,p}_{1,-s}(U_i)=\mathop{\arg\min}_{F_1(U_i)}{\sum_{t\neq s,t=1}^m(y_{it}-F_1(U_i)\hat{b}_{t1}^{k,p})^2}, i\in R_k.
\endn
Let $\breve F_1^{k,p} = (\breve F_1^{k,p}(U_1), \cdots,\breve F_1^{k,p}(U_n) )^\top$, where $\breve F_1^{k,p}(U_i) = \hat{F}^{k,p}_{1,-s}(U_i) I(i \in R_k) + \hat{F}^{k,p}_{1}(U_i) I(i \notin R_k)$, with $\hat{F}^{k,p}_{1}(U_i)$ given in Step 1.
Next, we estimate $F_2^0(U_i)$ by solving
\beginn
\tilde{F}^{k,p}_{2,-s}(U_i)=\mathop{\arg\min}_{F_2(U_i)}{\sum_{t\neq s,t=1}^m(y_{it}- \hat{F}^{k,p}_{1,-s}(U_i) \hat{b}_{t1}^{k,p} - F_2(U_i)\hat{b}_{t2}^{k,p})^2}, i\in R_k,
\endn
and let $\breve F_2^{k,p} = (\breve F_2^{k,p}(U_1), \cdots,\breve F_2^{k,p}(U_n) )^\top$, where $\breve F_2^{k,p}(U_i) = \tilde{F}^{k,p}_{2,-s}(U_i) I(i \in R_k) + \hat{F}^{k,p}_{2}(U_i) I(i \notin R_k)$, with $\hat{F}^{k,p}_{2}(U_i)$ given in Step 1. Recall that the OLS residuals are orthogonal to the regressors in linear regression, thus we regress $\tilde F_2^{k,p}$ on $\breve F_1^{k,p}$. As a result, the residual vector is orthogonal to $\breve F_1^{k,p}$ and the elements of residuals in $R_k$ are actually the estimators of $F_2^0(U_i)$ for $i \in R_k$, denoted by $\hat{F}^{k,p}_{2,-s}(U_i)$.
For $j=3,\cdots,p$, we  successively apply the similar procedure to estimate $F_j^0(U_i)$. By solving
\beginn
\tilde{F}^{k,p}_{j,-s}(U_i)=\mathop{\arg\min}_{F_j(U_i)}{\sum_{t\neq s,t=1}^m(y_{it}- \hat{F}^{k,p}_{1,-s}(U_i) \hat{b}_{t1}^{k,p} - \cdots - \hat{F}^{k,p}_{j-1,-s}(U_i) \hat{b}_{t,j-1}^{k,p} - F_j(U_i)\hat{b}_{tj}^{k,p}})^2, i\in R_k,
\endn
and define $\breve F_j^{k,p} = (\breve F_j^{k,p}(U_1), \cdots,\breve F_j^{k,p}(U_n) )^\top$, where $\breve F_j^{k,p}(U_i) = \tilde{F}^{k,p}_{j,-s}(U_i) I(i \in R_k) + \hat{F}^{k,p}_{j}(U_i) I(i \notin R_k)$. Regressing $\breve F_j^{k,p}$ on $\breve F_1^{k,p}, \cdots, \breve F_{j-1}^{k,p}$, then the residuals for $i \in R_k$ are actually the estimators in this step, denoted by $\hat{F}^{k,p}_{j,-s}(U_i)$. Hence, for each fixed $j$, we could compute the predicted value by $\hat y_{is} = \hat{F}_{-s}^{{k,p}^\top}(U_i)\hat{b}_s^{k,p}$, where $\hat{F}_{-s}^{k,p} = (\hat{F}^{k,p}_{1,-s},\cdots, \hat{F}^{k,p}_{j,-s})^\top$.

The average squared prediction error for $\Y_i=(y_{i1},...,y_{im})^\top$ is
\beginy
V_{i,k}^{[p]}=\frac{1}{m}\sum_{s=1}^{m}[y_{is}-\hat{F}_{-s}^{{k,p}^\top}(U_i)\hat{b}_s^{k,p}]^2,i\in R_k.
\endy

Finally, the averaged prediction error over all the elements with the working number of factors $p$ can be given by
\beginy
FTCV(p)=\frac{1}{n}\sum_{k=1}^{K}\sum_{i\in R_k}^{n}V_{i,k}^{[p]}.
\endy
Let $p_{max}$ be a fixed positive integer that is large enough such that $m>p_{max}>p_0$. The $FTCV$
estimator for the number of factors is given by
\beginy
\hat{p}_{FTCV}=\mathop{\arg\min}_{0\leqslant p\leqslant p_{max}}FTCV(p). \label{Model5}
\endy

The approach mentioned above is built upon the idea of double cross validation ($DCV$) \citep{zeng2019double}. However, due to the specific nature of our semiparametric factor models, the DCV method may not be applied without incorporating information from $U$. Additionally, our proposed FTCV method not only allows for the selection of the number of common functional factors, but also offers an easily implemented procedure by considering the uncorrelatedness of functional factors.

\section{Simulation studies} \label{sec5}
\par In this section, we conduct experimental studies to evaluate the finite sample performance of the proposed FTCV method. In order to make comparisons, representative approaches such as $IC_\p$ (eigen-structure based) and $Ladle$ (IC-based) are included in the studies. In addition, we consider both leave-one-out $FTCV$ ($FTCV_1$) and $10$-fold $FTCV$ ($FTCV_{10}$) in the first step.
\begin{figure}[htb]
\centering
    \begin{minipage}[t]{0.34\linewidth}
        \centering
        \includegraphics[scale=0.365]{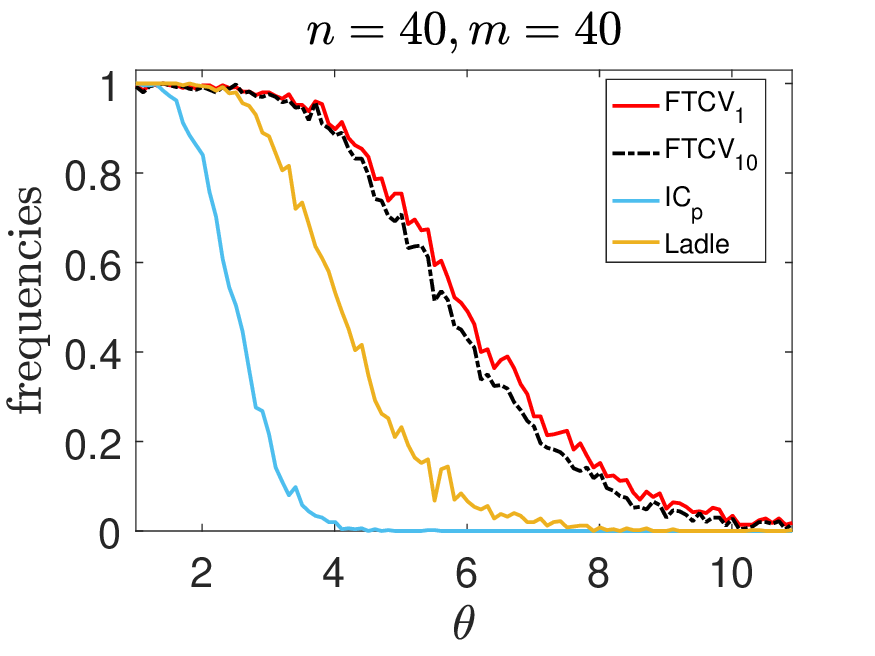}
        \vspace{0.2cm}
        \includegraphics[scale=0.365]{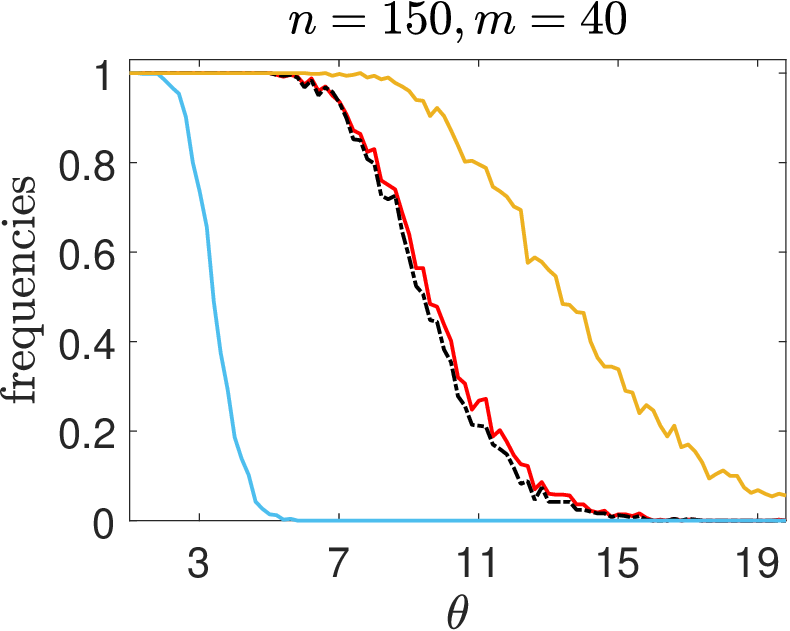}
        \vspace{0.2cm}
        \includegraphics[scale=0.365]{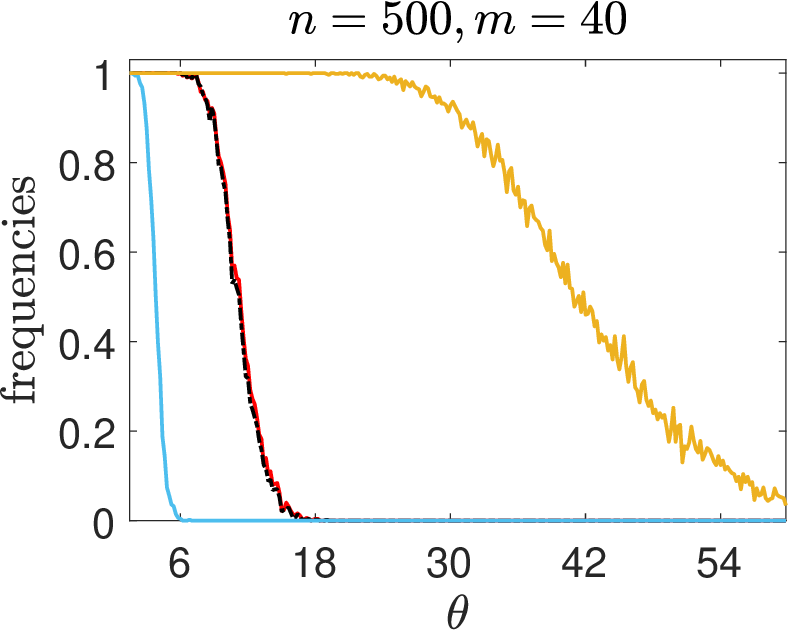}
    \end{minipage}%
    \begin{minipage}[t]{0.34\linewidth}
    \centering
        \includegraphics[scale=0.365]{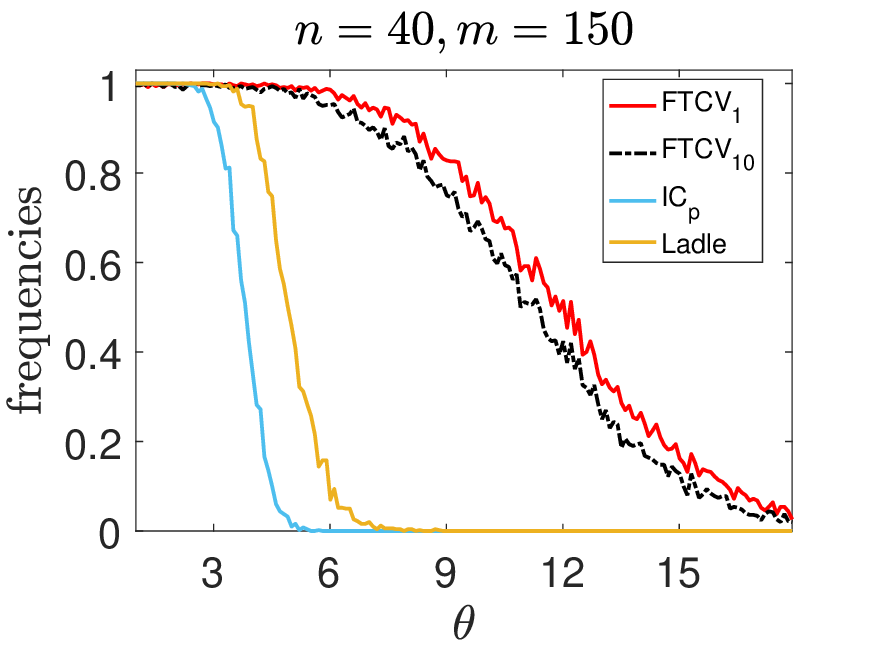}
        \vspace{0.2cm}
        \includegraphics[scale=0.365]{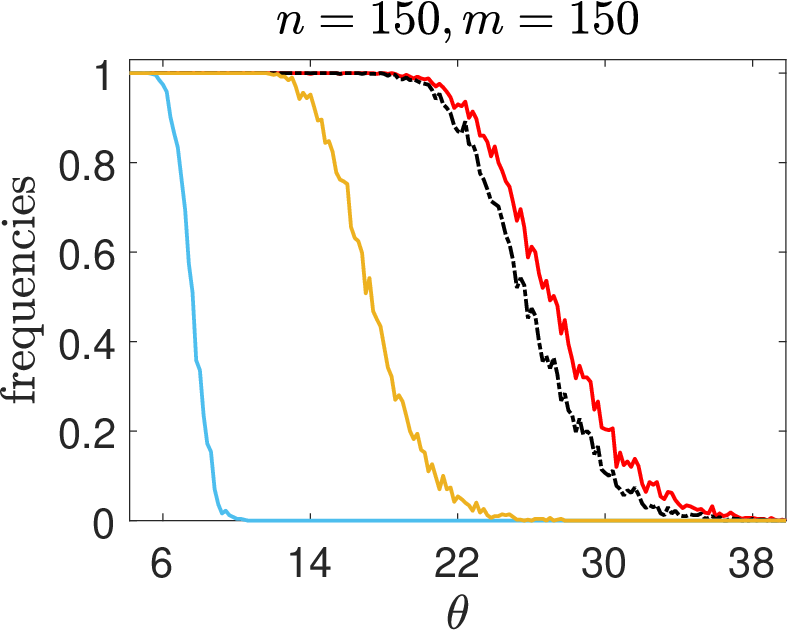}
        \vspace{0.2cm}
        \includegraphics[scale=0.365]{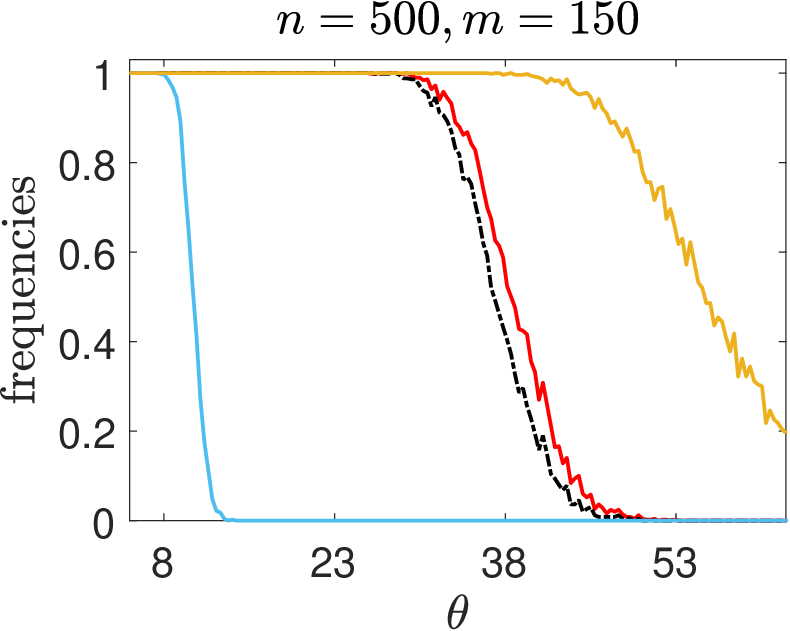}
    \end{minipage}%
    \begin{minipage}[t]{0.34\linewidth}
        \centering
        \includegraphics[scale=0.365]{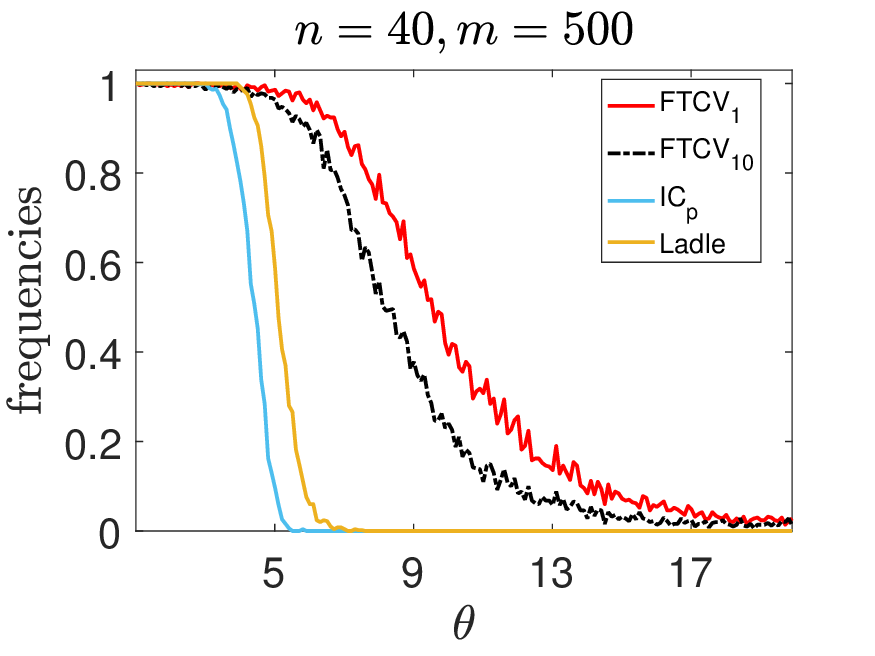}
        \vspace{0.2cm}
        \includegraphics[scale=0.365]{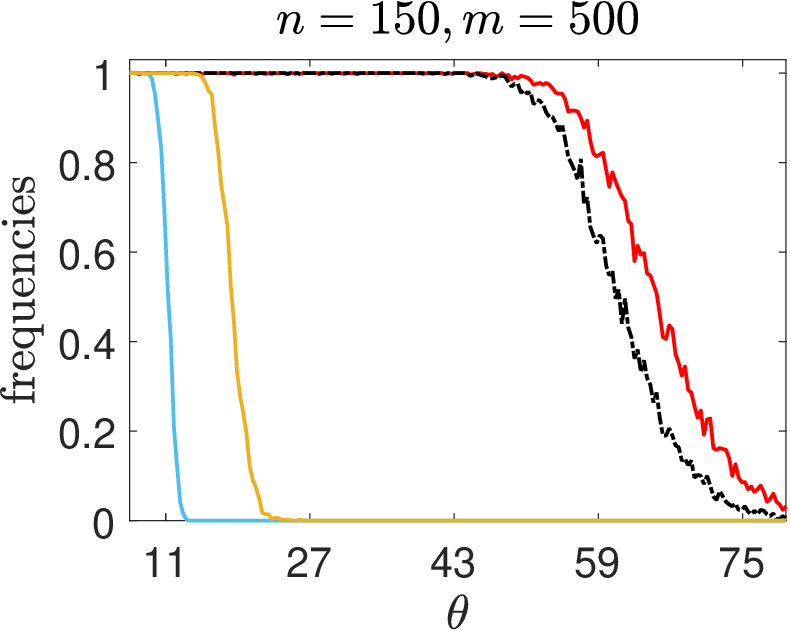}
        \vspace{0.2cm}
        \includegraphics[scale=0.365]{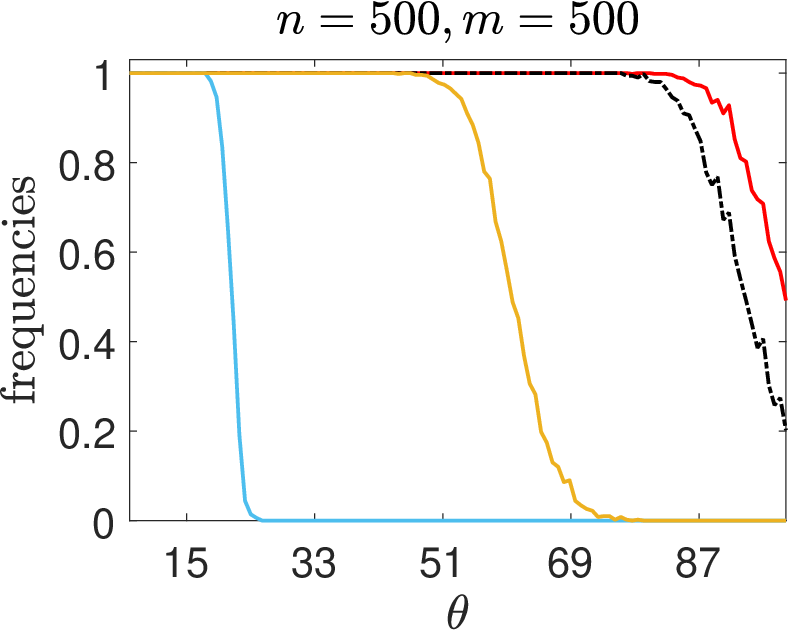}
    \end{minipage}%
\centering
\caption{\scriptsize{The relative frequencies of correctly selecting the number of factors for model \eqref{Models} under Scenario 1 with independent Gaussian errors (E1). In each panel, we use red curve for $FTCV_1$, black dash dotted curve for $FTCV_{10}$, blue curve for $IC_\p$, and yellow curve for $Ladle$. }}
\label{fig1}
\end{figure}

We simulate data from the following semiparametric model:
\beginy \label{Models}
y_{is}=\sum_{j=1}^{p_0}F_j(U_i)b_{sj}+\sqrt{\theta}e_{is},\ \  i=1,...,n, \ \ s=1,...,m.
\endy
where $b_{sj} (j=1,\cdots, p_0)$ are constant coefficients and $e_{is}$ are idiosyncratic errors.
For all simulation studies in this section, $\theta$ is introduced to reflect the effect of the noise level. In order to demonstrate the robustness of our method, we consider two illustrative scenarios as follows:
\begin{figure}[ht]
\centering
    \begin{minipage}[t]{0.34\linewidth}
        \centering
        \includegraphics[scale=0.365]{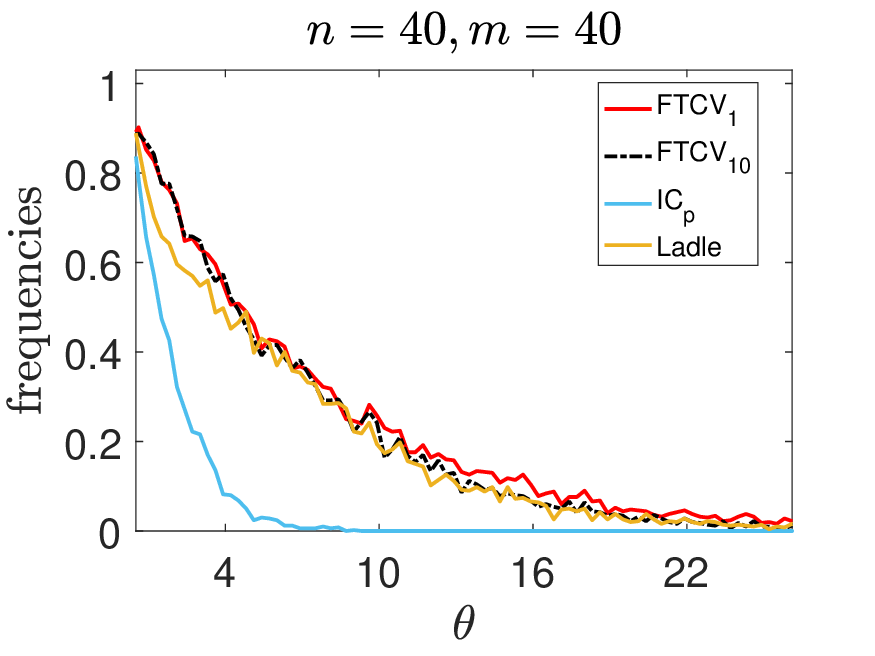}
        \vspace{0.2cm}
        \includegraphics[scale=0.365]{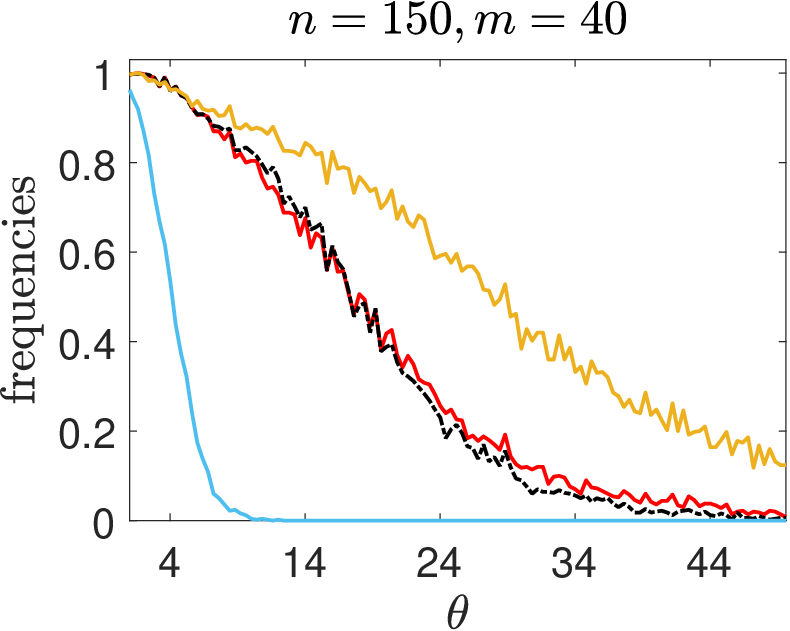}
        \vspace{0.2cm}
        \includegraphics[scale=0.365]{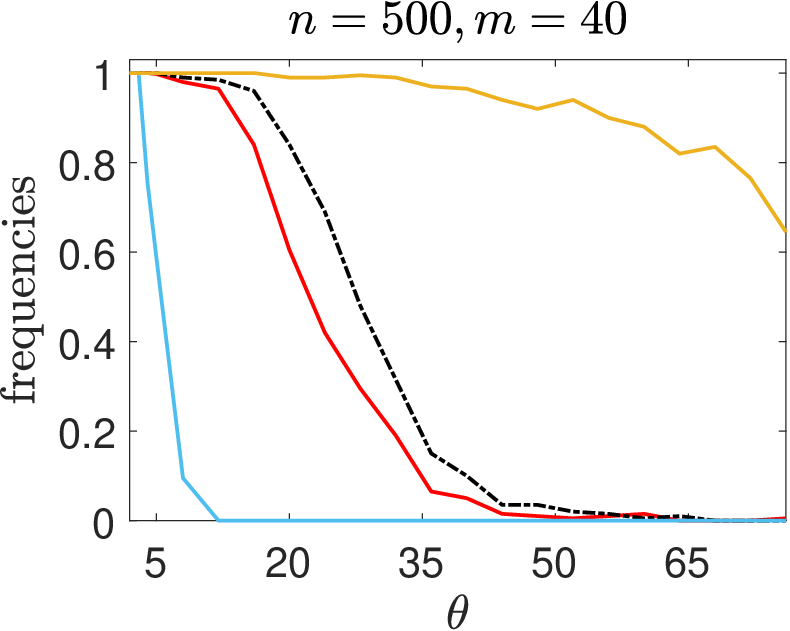}
    \end{minipage}%
    \begin{minipage}[t]{0.34\linewidth}
        \centering
        \includegraphics[scale=0.365]{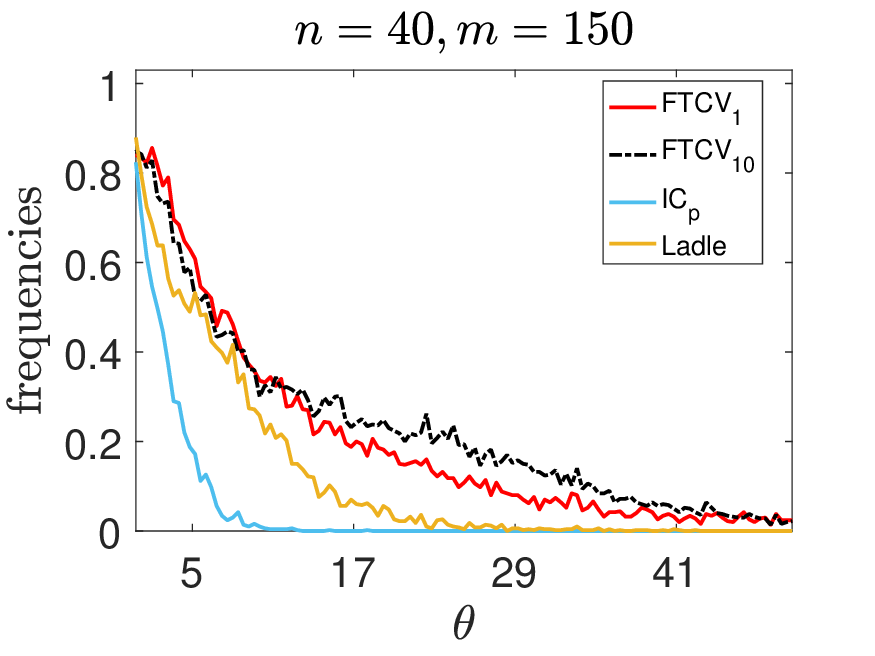}
        \vspace{0.2cm}
        \includegraphics[scale=0.365]{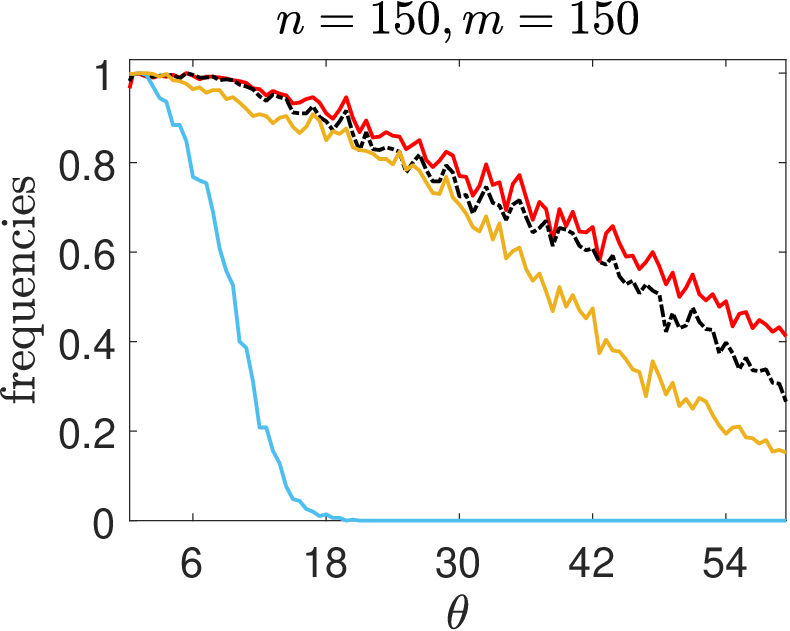}
        \vspace{0.2cm}
       \includegraphics[scale=0.365]{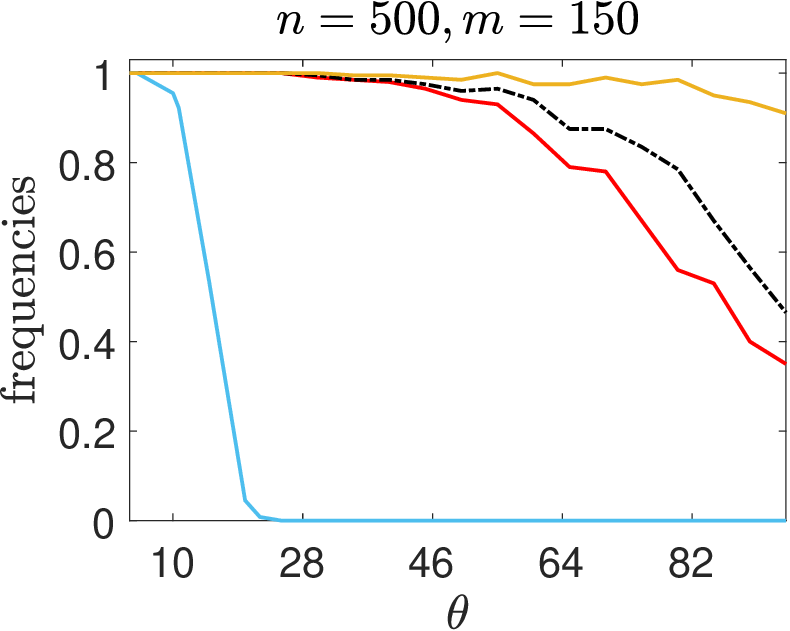}
    \end{minipage}%
    \begin{minipage}[t]{0.34\linewidth}
        \centering
        \includegraphics[scale=0.365]{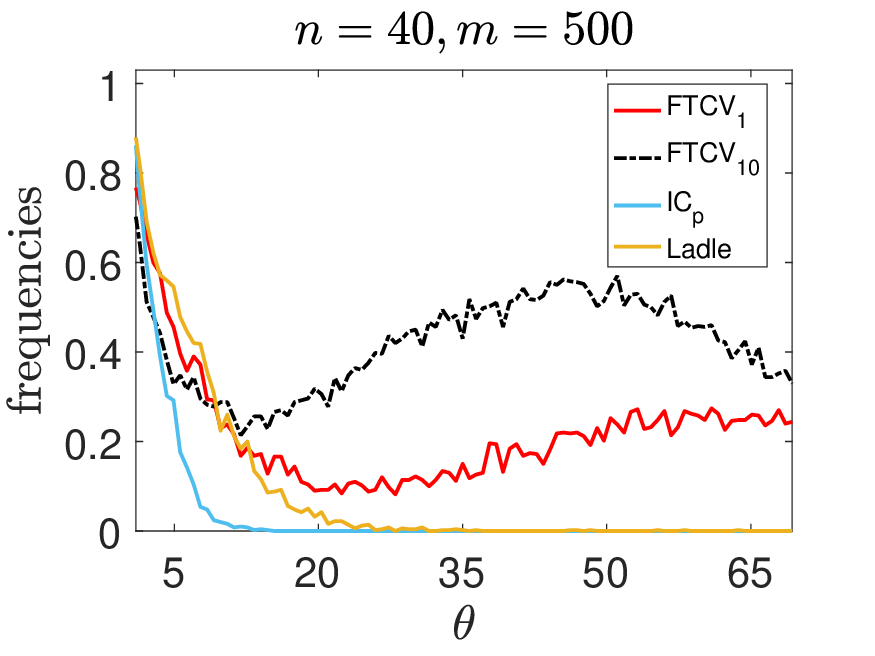}
        \vspace{0.2cm}
        \includegraphics[scale=0.365]{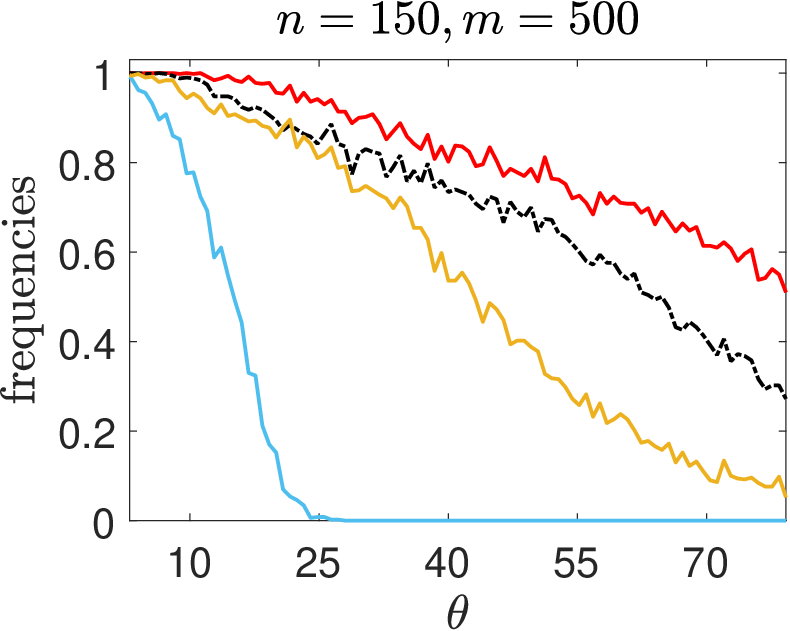}
        \vspace{0.2cm}
        \includegraphics[scale=0.365]{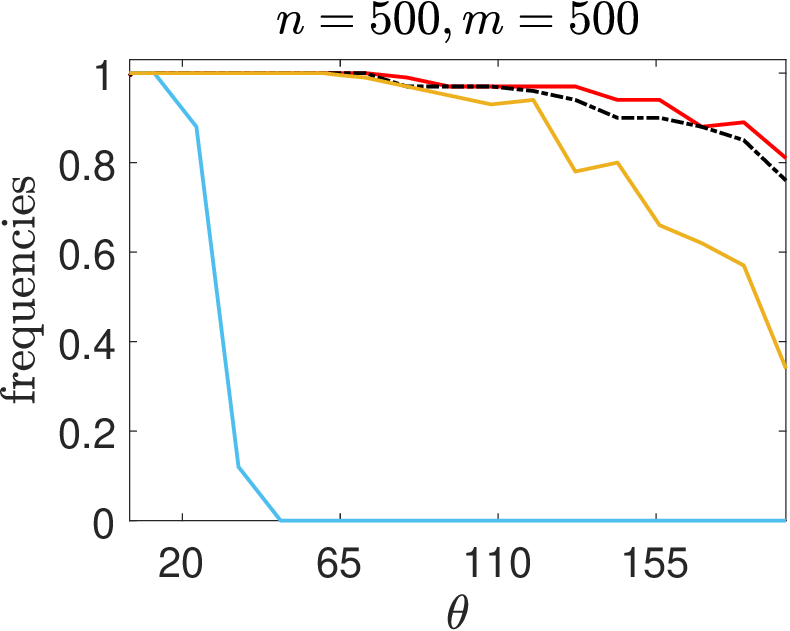}
    \end{minipage}%

\centering
\caption{\scriptsize{The relative frequencies of correctly selecting the number of factors for model \eqref{Models} under Scenario 2 with independent Gaussian errors (E1). In each panel, we use red curve for $FTCV_1$, black dash dotted curve for $FTCV_{10}$, blue curve for $IC_\p$, and yellow curve for $Ladle$.}}
\label{fig2}
\end{figure}

\par \textbf{Scenario 1:} Let the true number of factors $p_0=2$, $b_{s1}, b_{s2} \sim N(0,1)$ and $y_{is}$ involves two common factors defined by
\beginy
F_1(U)=\cos(2\pi U),\ \
F_2(U)=\sin(2\pi U),
\endy
where $U\sim Uniform(-1,1)$.
\par \textbf{Scenario 2:} Let the true number of factors $p_0=3$, $b_{s1}, b_{s2}, b_{s3} \sim N(0,1)$, a special setting was designed as follows:
\beginy
F_1(U)=U,\ \
F_2(U)=U^2-1, \ \
F_3(U)=0.4(U^4-6U^2+3),
\endy
where $U\sim N(0,1)$. However, different from traditional model framework, we allow one of the three coefficients for each $y_{is}$ to be randomly assigned the value $0$. Such specifications suggest that our entire model actually involves $3$ common factors but only $2$ factors make contributions to each $y_{is}$.
\par For the idiosyncratic errors $e_{is}$, we consider the following settings
\begin{description}
  \item[(E1)] independent Gaussian: $e_{is}\sim N(0,1)$;
  \item[(E2)] heteroskedastic: $e_{is} \sim N(0,\sigma_s^2)$, $\sigma_s^2=1$ if $s$ is odd, and $\sigma_s^2=2$ if $s$ is even;
  \item[(E3)] cross correlated: $e_{i,\cdot} = (e_{i1},\cdots,e_{im})^\top$, $e_{i,\cdot} \sim N(\mathbf{0}, \bm\Sigma)$, where $\bm\Sigma$ is a $m\times m$ covariance matrix whose $(k, l)$th element is $\bm\Sigma_{kl} = 0.5^{|k-l|}$.
\end{description}
\begin{figure}[ht]
\centering
    \begin{minipage}[t]{0.34\linewidth}
        \centering
        \includegraphics[scale=0.365]{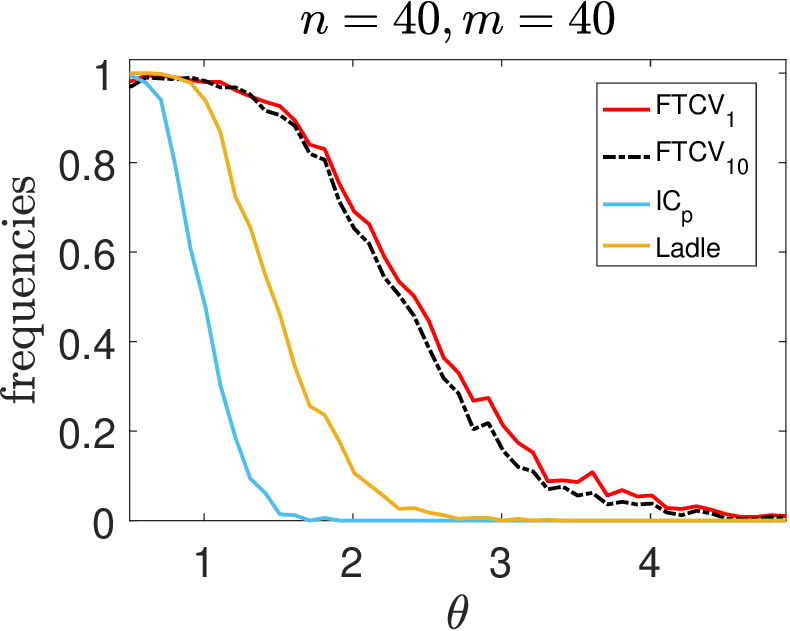}
        \vspace{0.2cm}
       \includegraphics[scale=0.365]{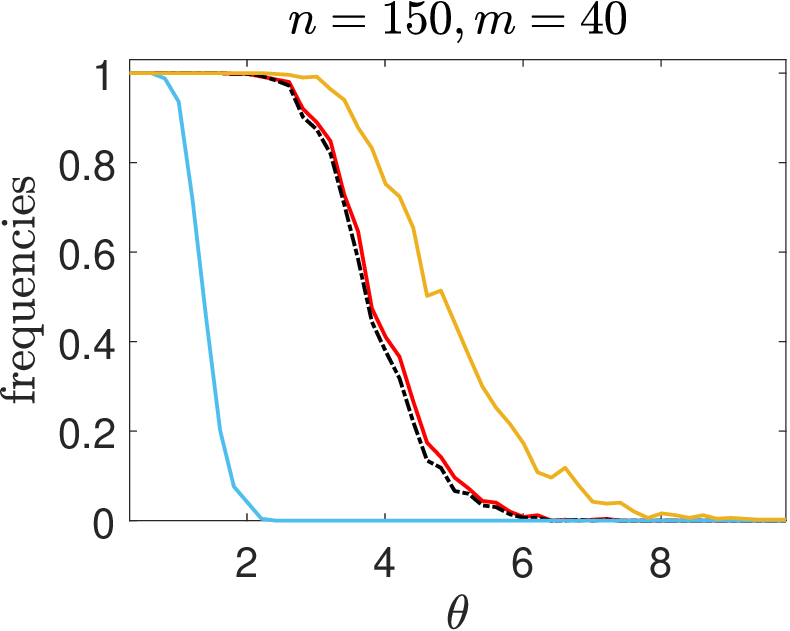}
        \vspace{0.2cm}
        \includegraphics[scale=0.365]{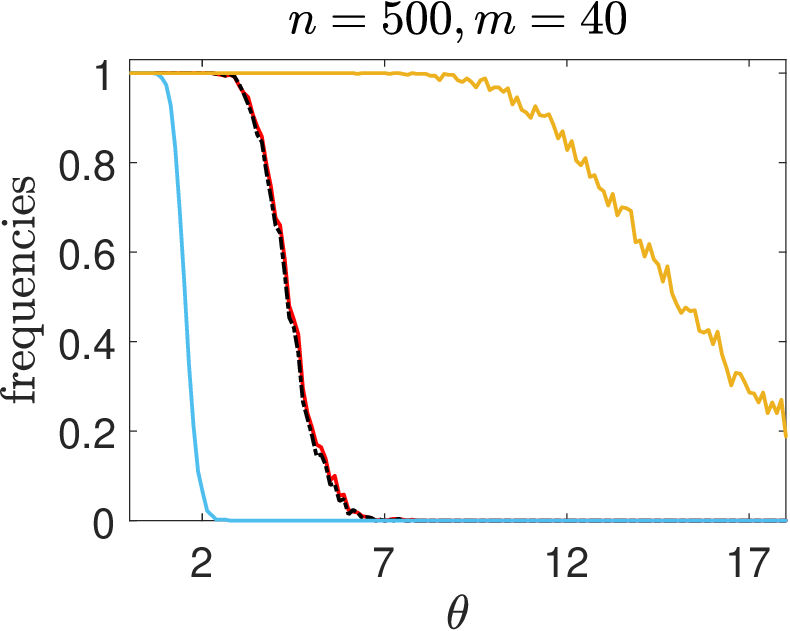}
    \end{minipage}%
    \begin{minipage}[t]{0.34\linewidth}
        \centering
        \includegraphics[scale=0.365]{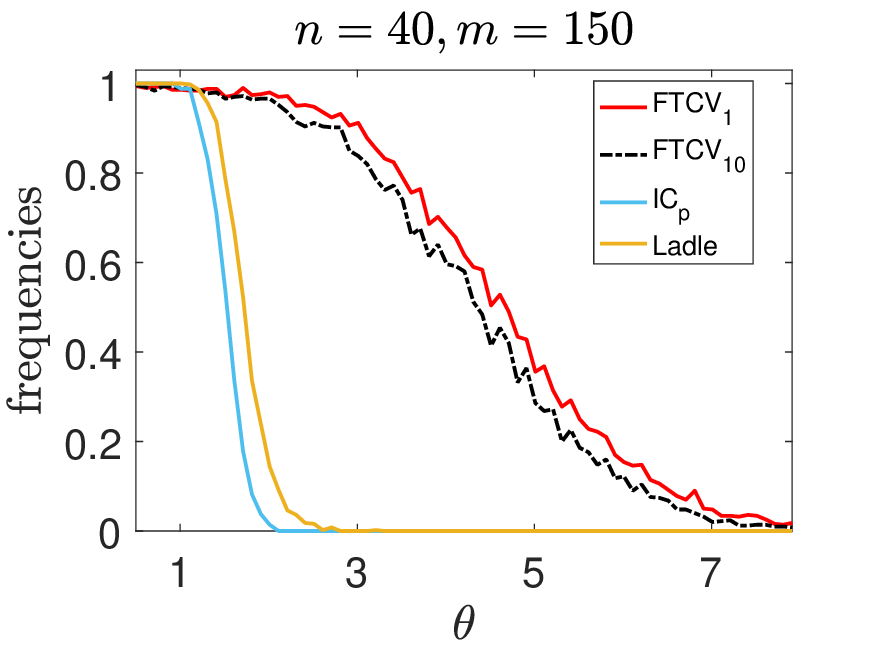}
        \vspace{0.2cm}
      \includegraphics[scale=0.365]{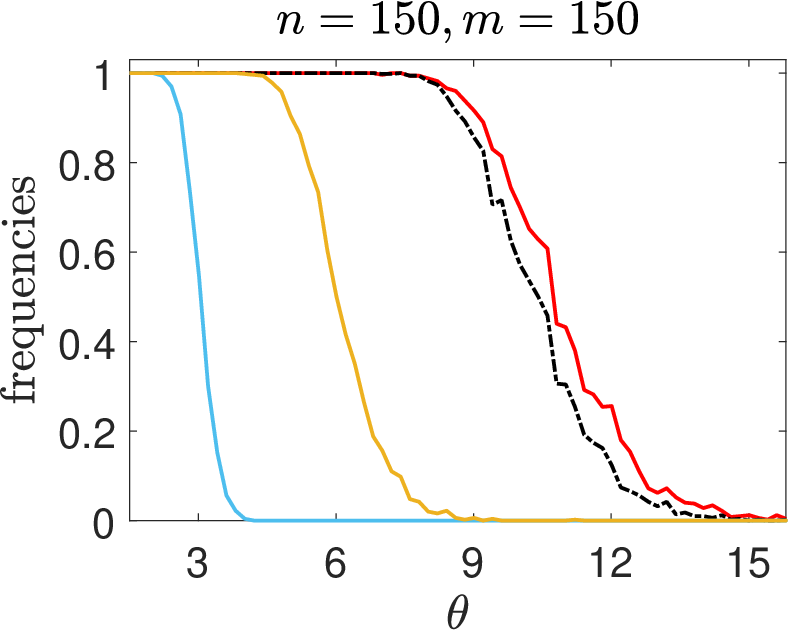}
        \vspace{0.2cm}
      \includegraphics[scale=0.365]{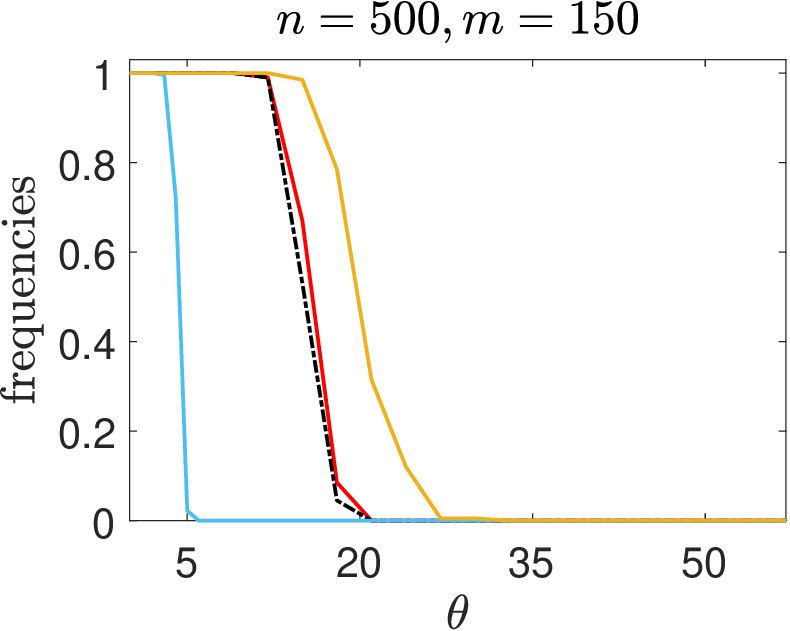}
    \end{minipage}%
    \begin{minipage}[t]{0.34\linewidth}
        \centering
        \includegraphics[scale=0.365]{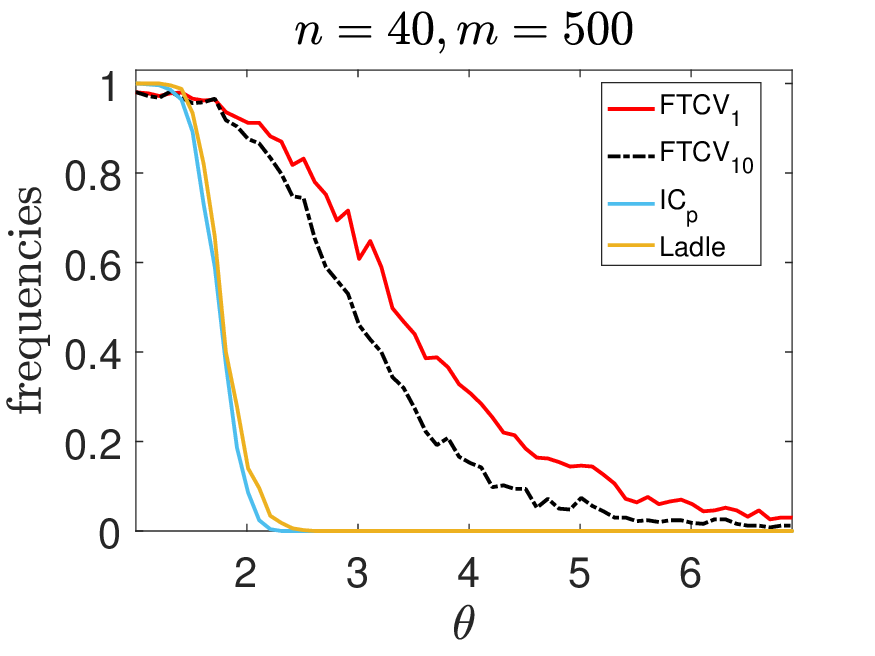}
        \vspace{0.2cm}
        \includegraphics[scale=0.365]{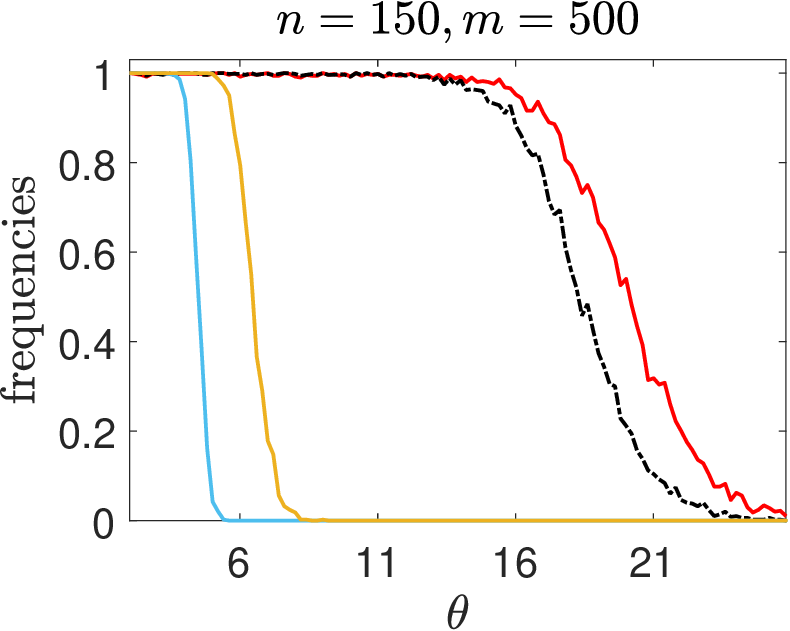}
        \vspace{0.2cm}
        \includegraphics[scale=0.365]{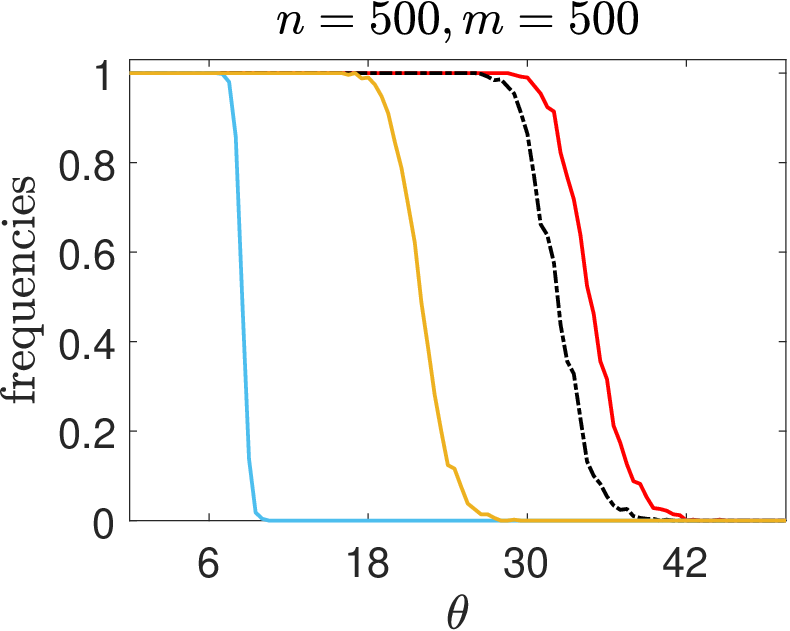}
    \end{minipage}%

\centering
\caption{\scriptsize{The relative frequencies of correctly selecting the number of factors for model \eqref{Models} under Scenario 1 with heteroskedastic errors (E2).  In each panel, we use red curve for $FTCV_1$, black dash dotted curve for $FTCV_{10}$, blue curve for $IC_\p$, and yellow curve for $Ladle$. }}
\label{fig3}
\end{figure}

\par
\begin{figure}[ht]
\centering
    \begin{minipage}[t]{0.34\linewidth}
        \centering
        \includegraphics[scale=0.365]{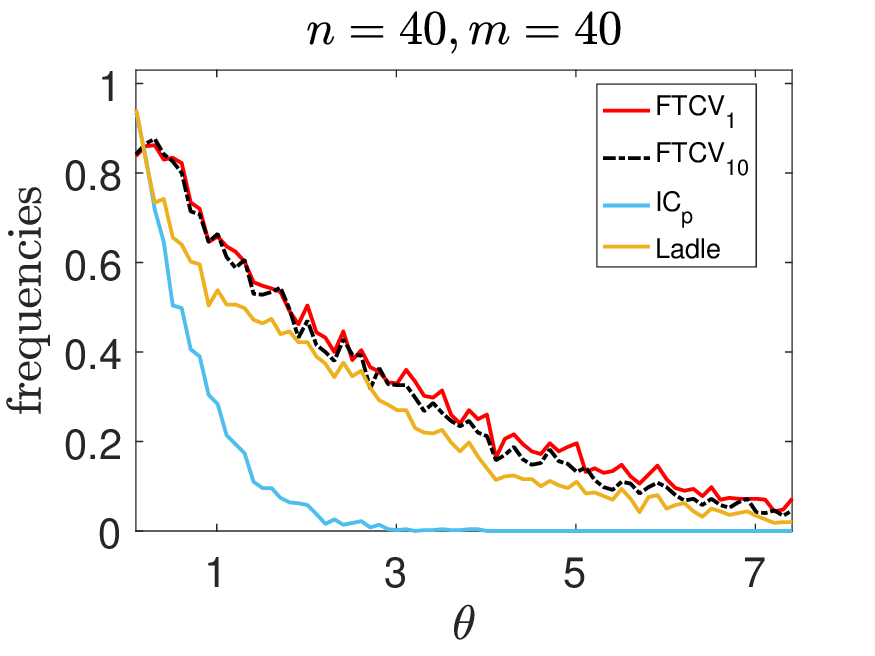}
        \vspace{0.2cm}
        \includegraphics[scale=0.365]{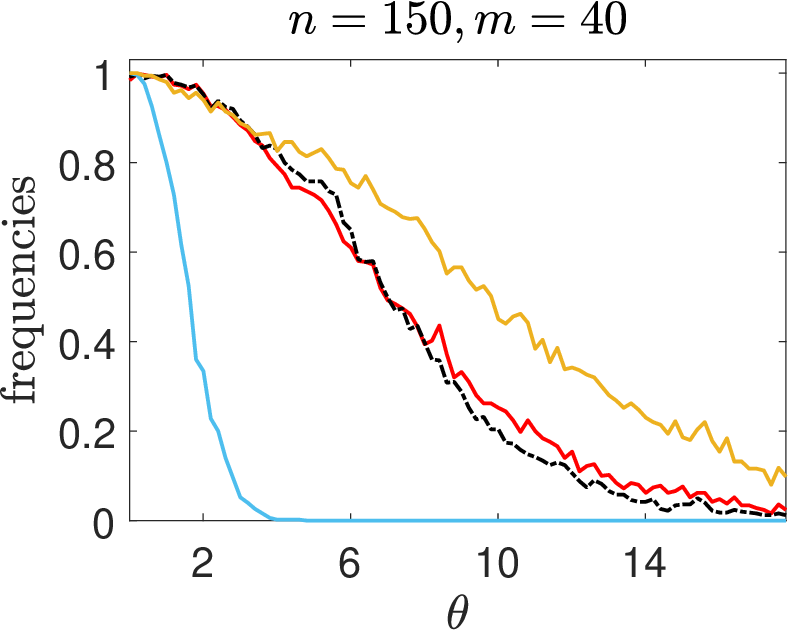}
       \vspace{0.2cm}
       \includegraphics[scale=0.365]{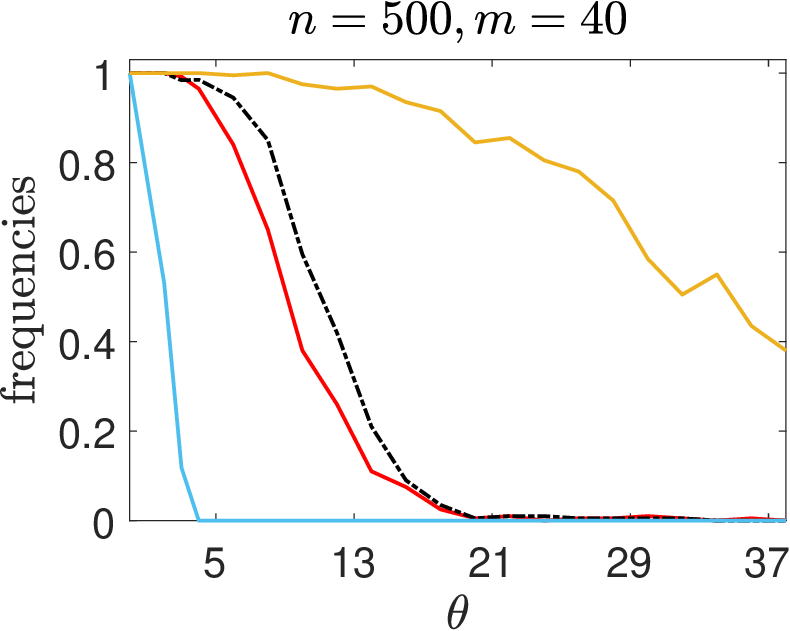}
    \end{minipage}%
    \begin{minipage}[t]{0.34\linewidth}
        \centering
       \includegraphics[scale=0.365]{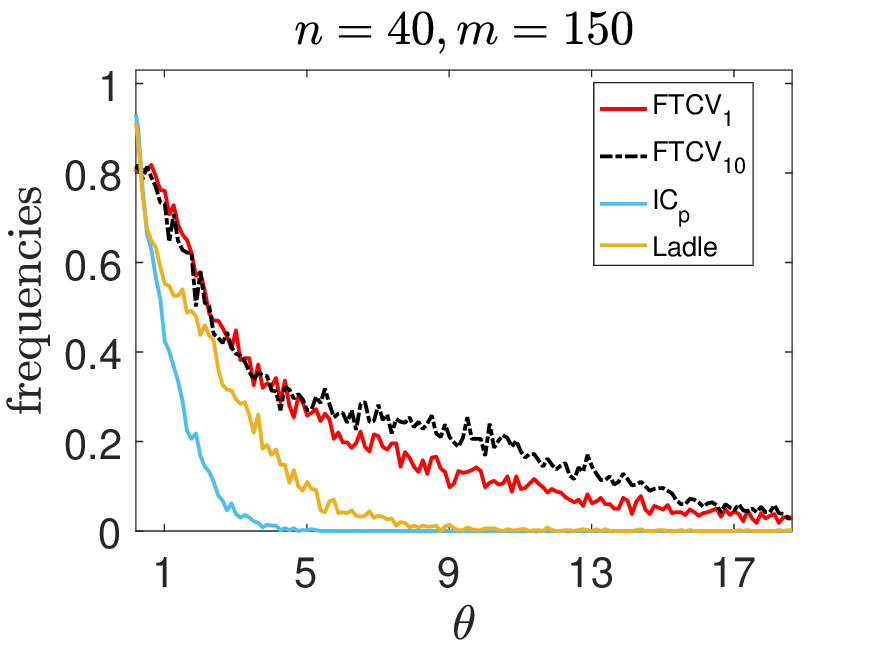}
        \vspace{0.2cm}
      \includegraphics[scale=0.365]{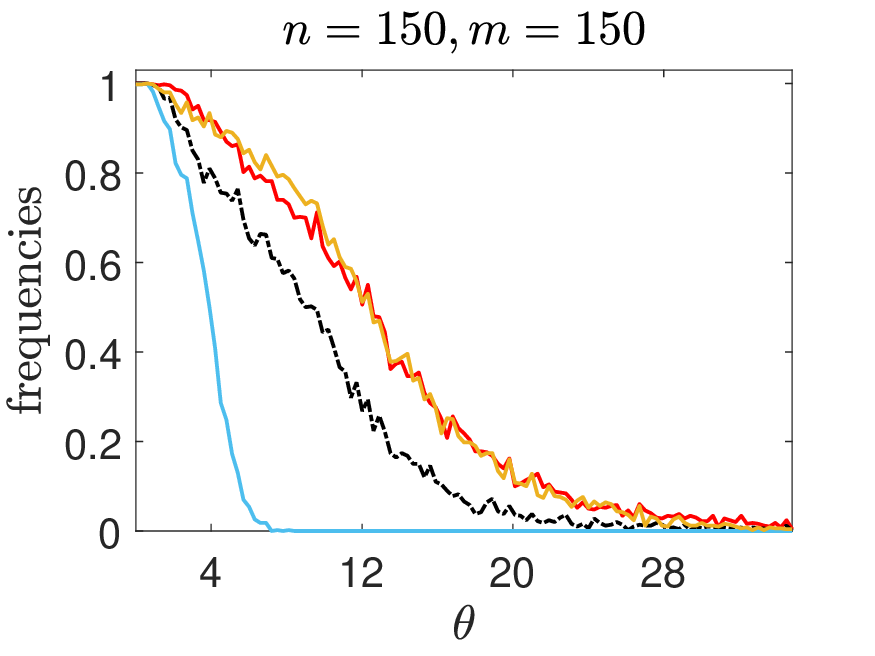}
        \vspace{0.2cm}
      \includegraphics[scale=0.365]{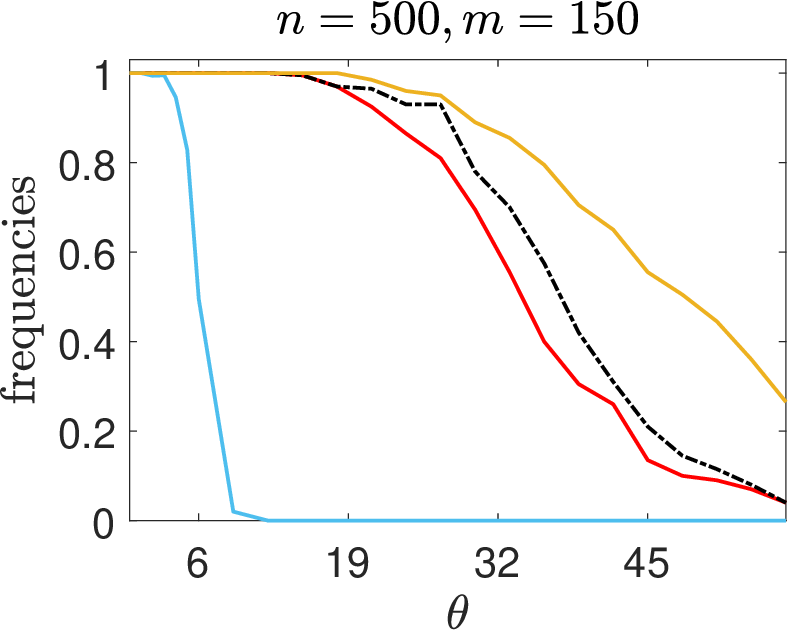}
    \end{minipage}%
    \begin{minipage}[t]{0.34\linewidth}
        \centering
        \includegraphics[scale=0.365]{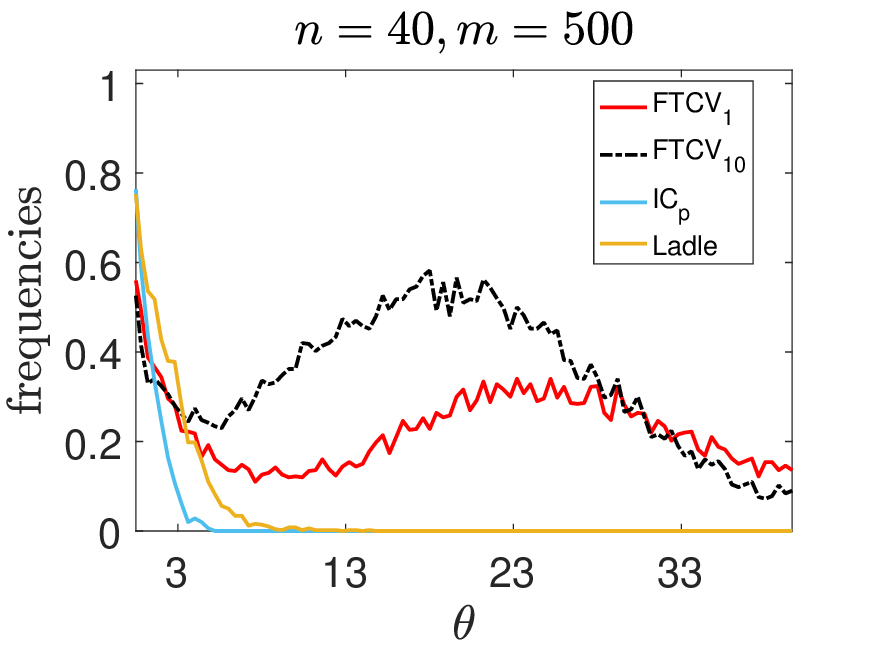}
        \vspace{0.2cm}
       \includegraphics[scale=0.365]{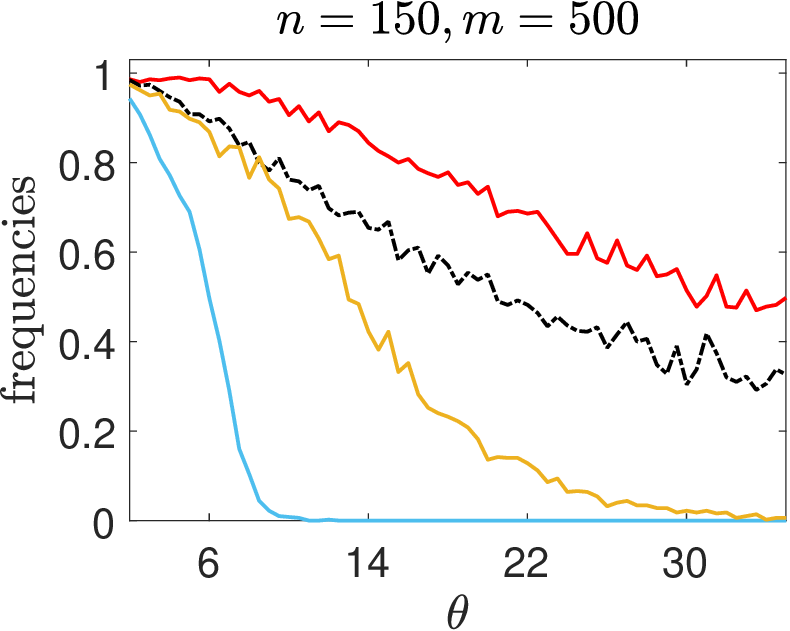}
        \vspace{0.2cm}
        \includegraphics[scale=0.365]{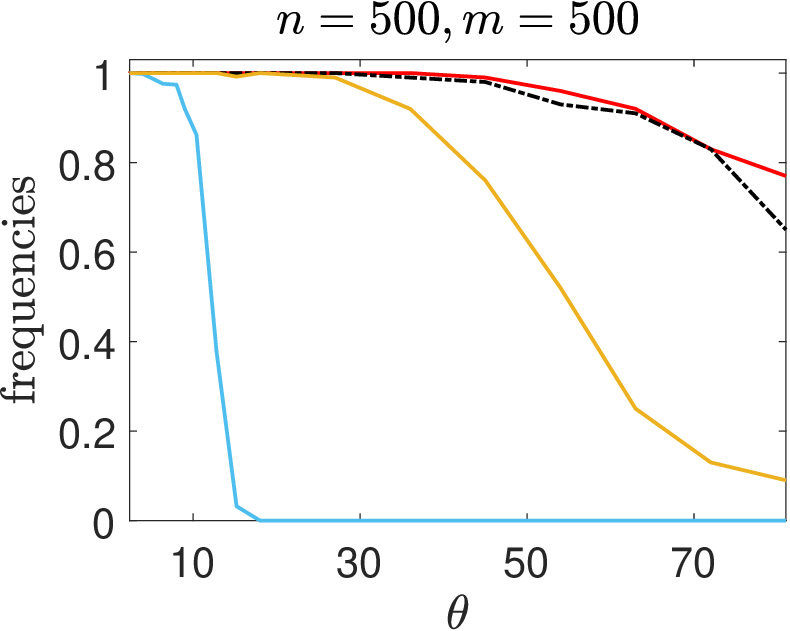}
    \end{minipage}%

\centering
\caption{\scriptsize{The relative frequencies of correctly selecting the number of factors for model \eqref{Models} under Scenario 2 with heteroskedastic errors (E2).  In each panel, we use red curve for $FTCV_1$, black dash dotted curve for $FTCV_{10}$, blue curve for $IC_\p$, and yellow curve for $Ladle$. }}
\label{fig4}
\end{figure}

\begin{figure}[ht]
\centering
    \begin{minipage}[t]{0.34\linewidth}
        \centering
        \includegraphics[scale=0.365]{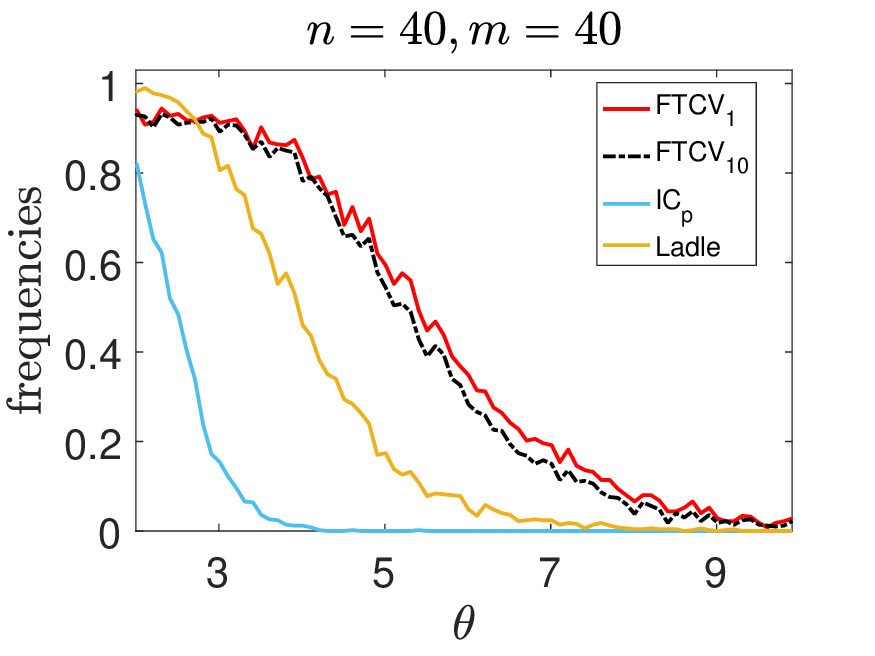}
        \vspace{0.2cm}
        \includegraphics[scale=0.365]{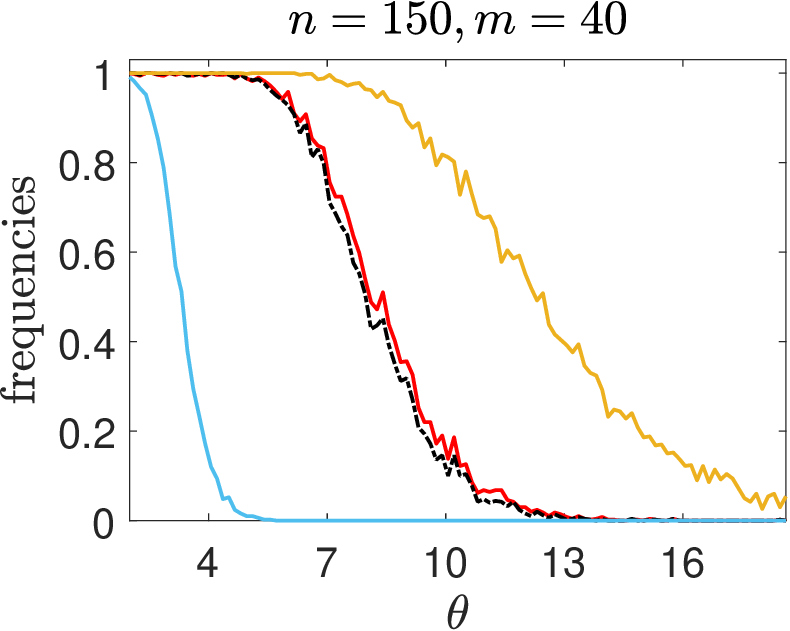}
        \vspace{0.2cm}
       \includegraphics[scale=0.365]{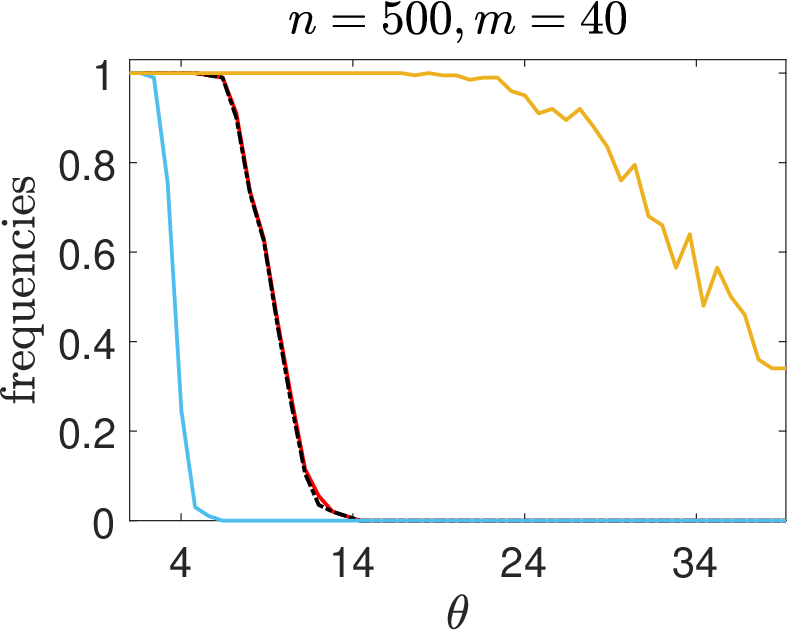}
    \end{minipage}%
    \begin{minipage}[t]{0.34\linewidth}
        \centering
        \includegraphics[scale=0.365]{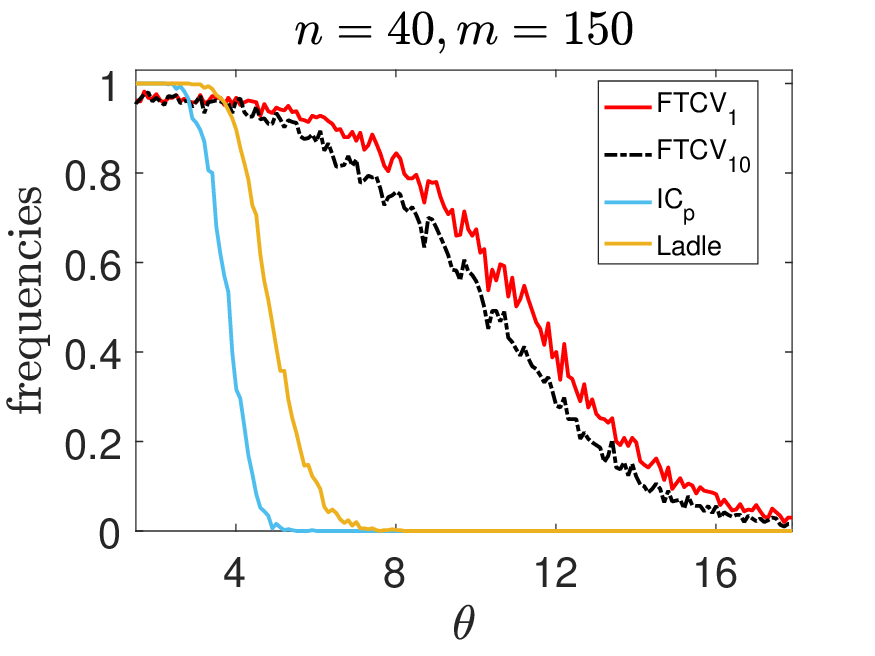}
        \vspace{0.2cm}
      \includegraphics[scale=0.365]{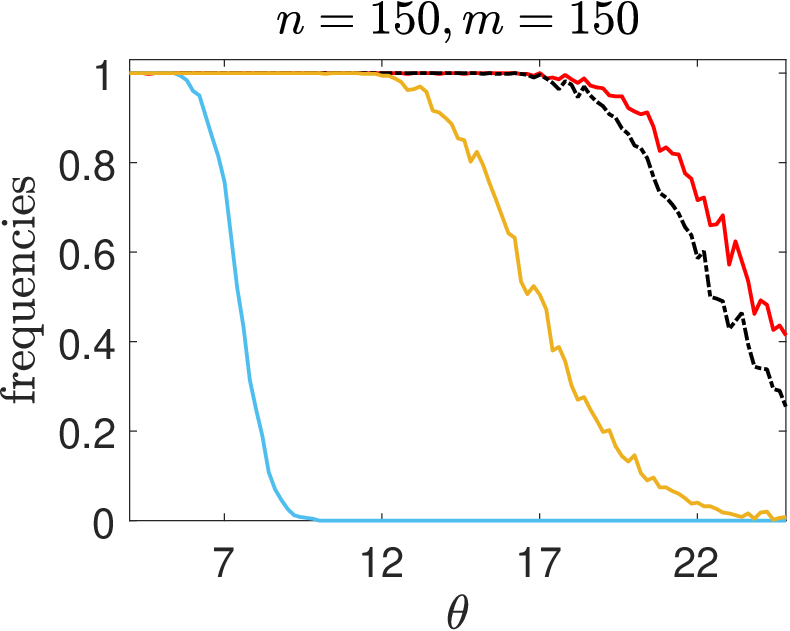}
        \vspace{0.2cm}
      \includegraphics[scale=0.365]{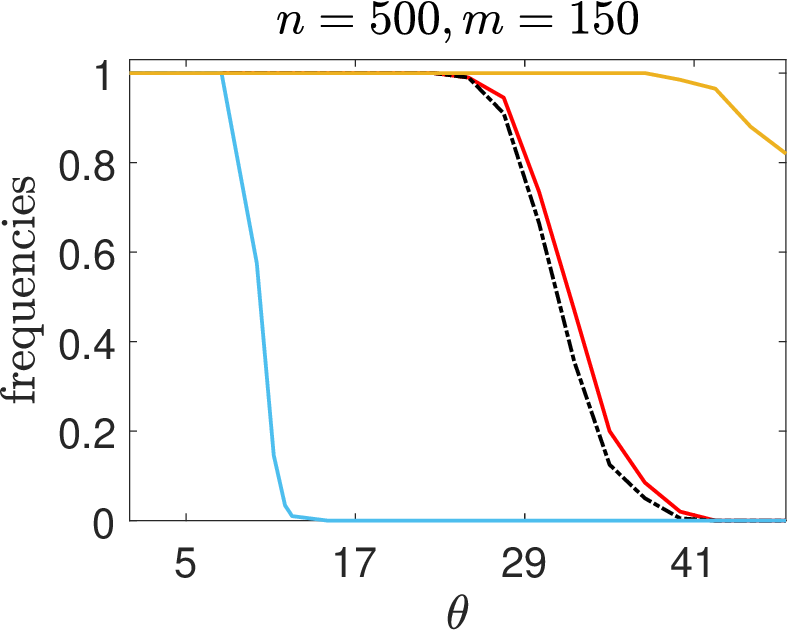}
    \end{minipage}%
    \begin{minipage}[t]{0.34\linewidth}
        \centering
        \includegraphics[scale=0.365]{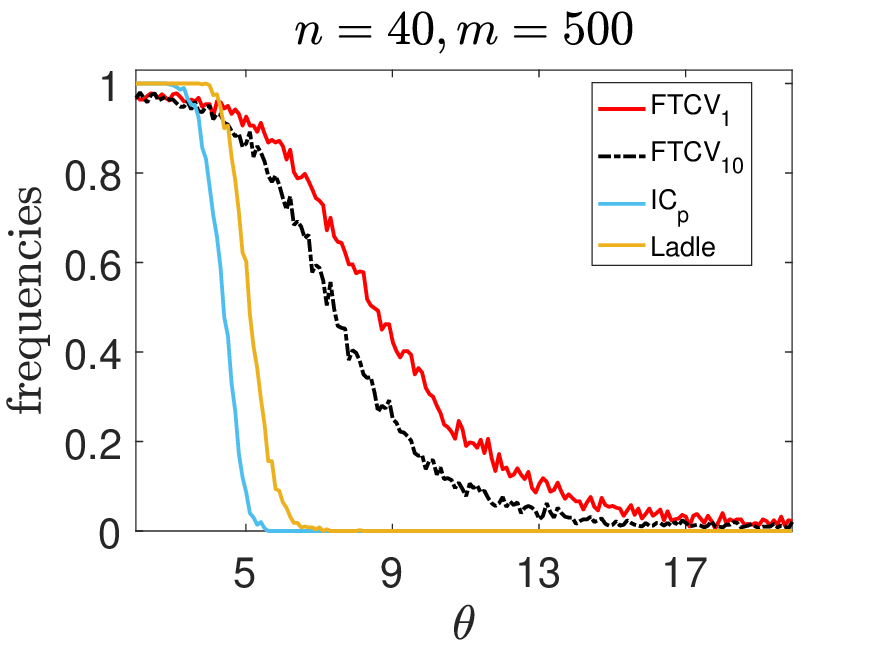}
        \vspace{0.2cm}
        \includegraphics[scale=0.365]{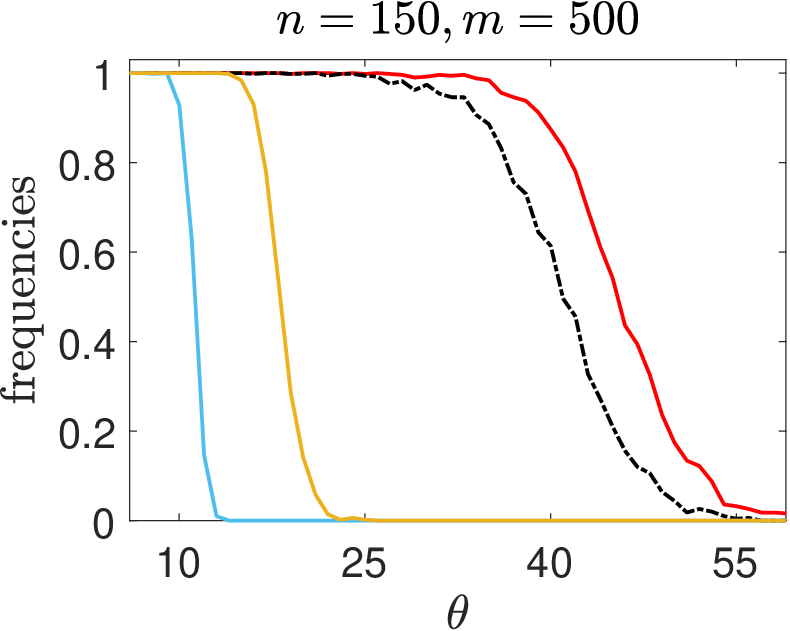}
        \vspace{0.2cm}
        \includegraphics[scale=0.365]{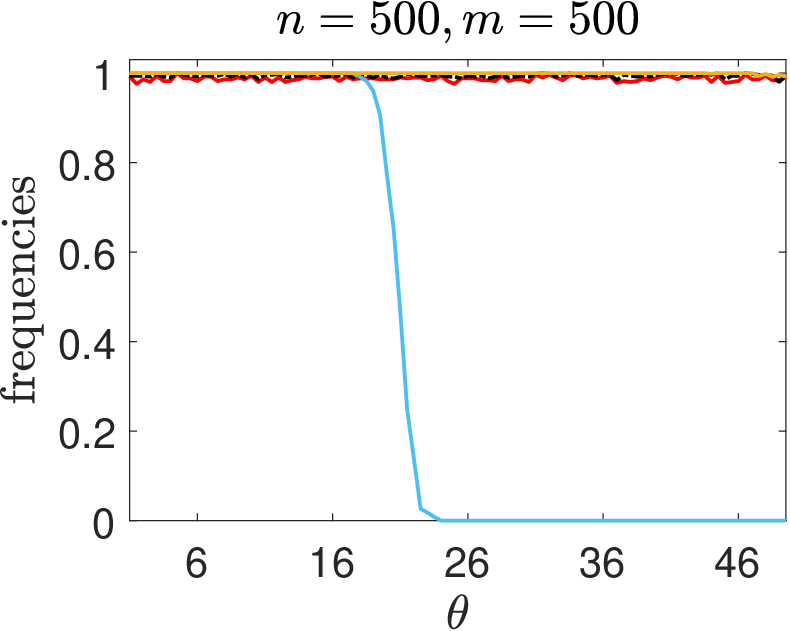}
    \end{minipage}%

\centering
\caption{\scriptsize{The relative frequencies of correctly selecting the number of factors for model \eqref{Models} under Scenario 1 with cross correlated errors (E3).  In each panel, we use red curve for $FTCV_1$, black dash dotted curve for $FTCV_{10}$, blue curve for $IC_\p$, and yellow curve for $Ladle$. }}
\label{fig5}
\end{figure}

\begin{figure}[ht]
\centering
    \begin{minipage}[t]{0.34\linewidth}
        \centering
        \includegraphics[scale=0.365]{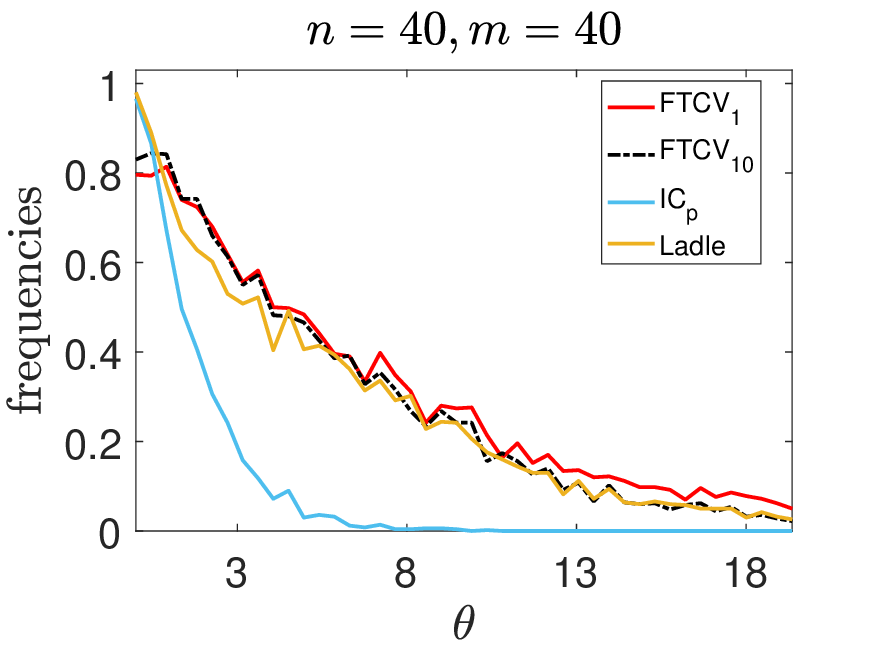}
        \vspace{0.2cm}
       \includegraphics[scale=0.365]{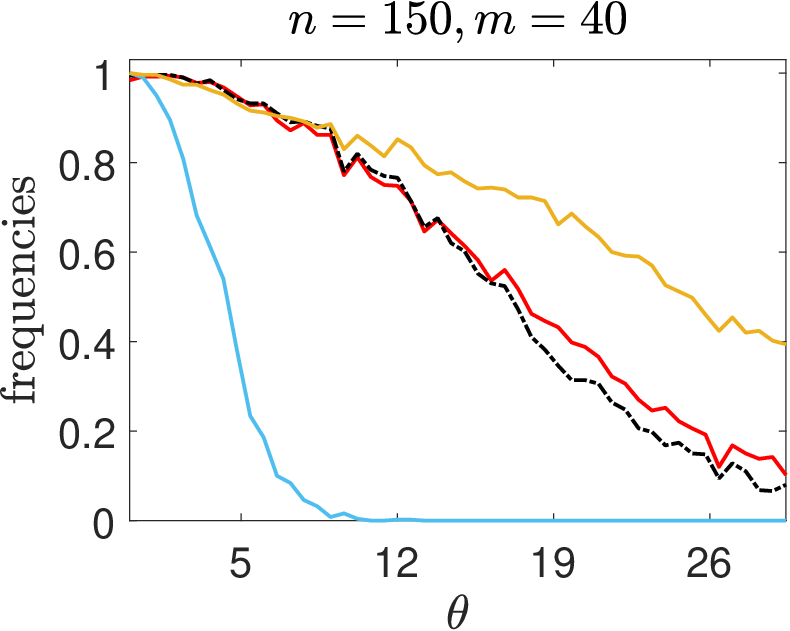}
       \vspace{0.2cm}
       \includegraphics[scale=0.365]{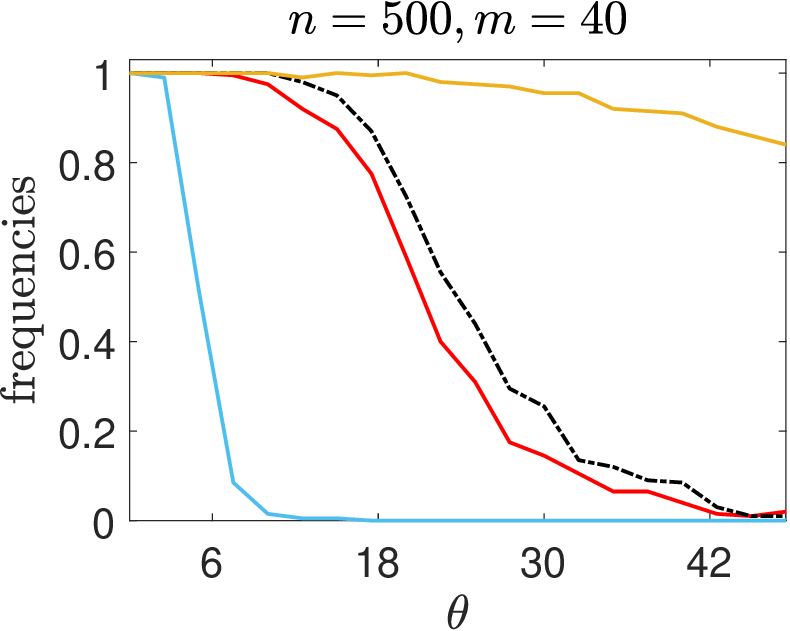}
    \end{minipage}%
    \begin{minipage}[t]{0.34\linewidth}
        \centering
       \includegraphics[scale=0.365]{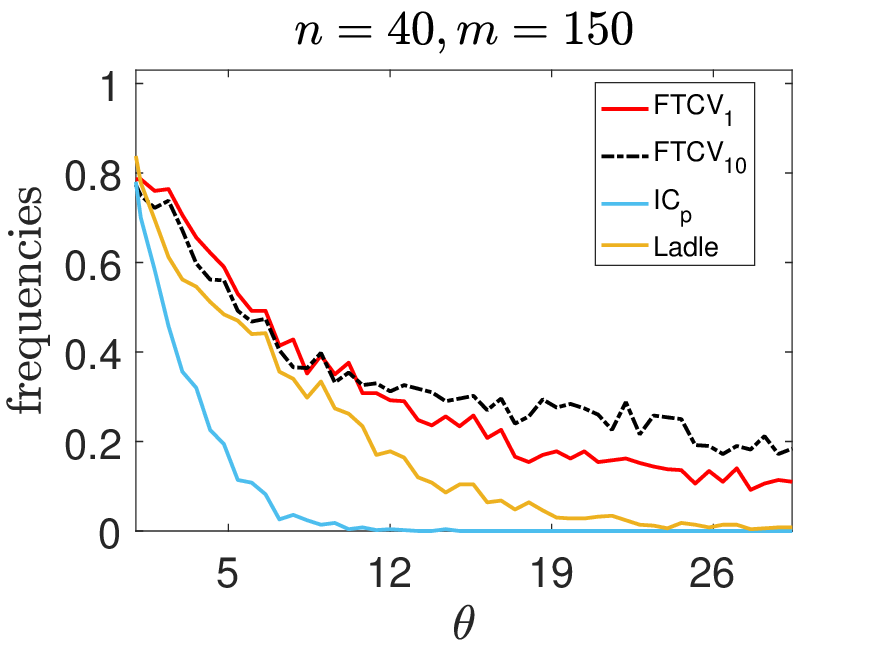}
        \vspace{0.2cm}
       \includegraphics[scale=0.365]{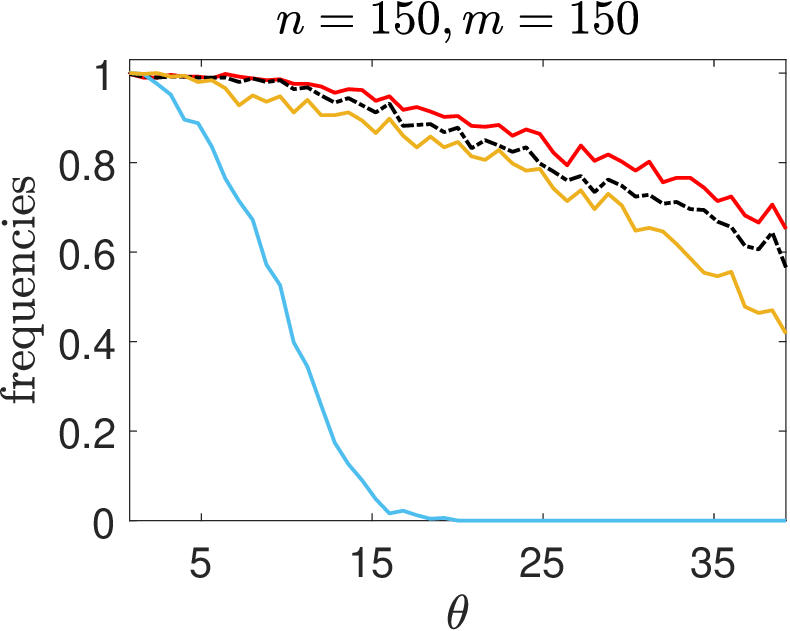}
        \vspace{0.2cm}
      \includegraphics[scale=0.365]{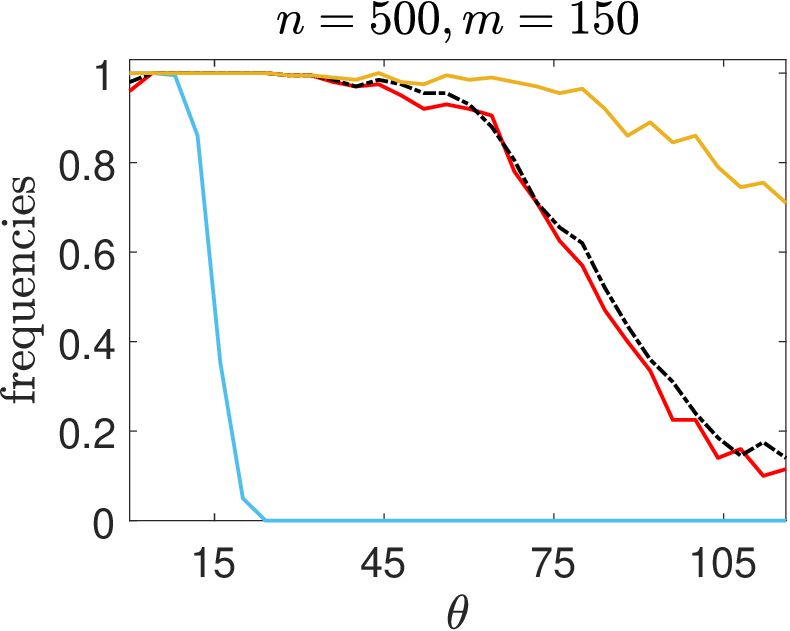}
    \end{minipage}%
    \begin{minipage}[t]{0.34\linewidth}
        \centering
        \includegraphics[scale=0.365]{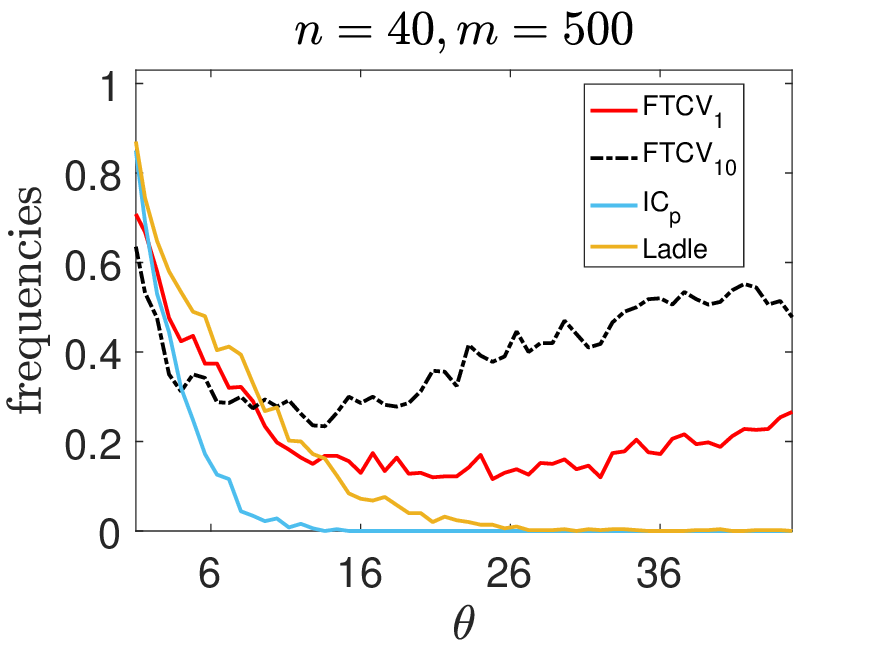}
        \vspace{0.2cm}
        \includegraphics[scale=0.365]{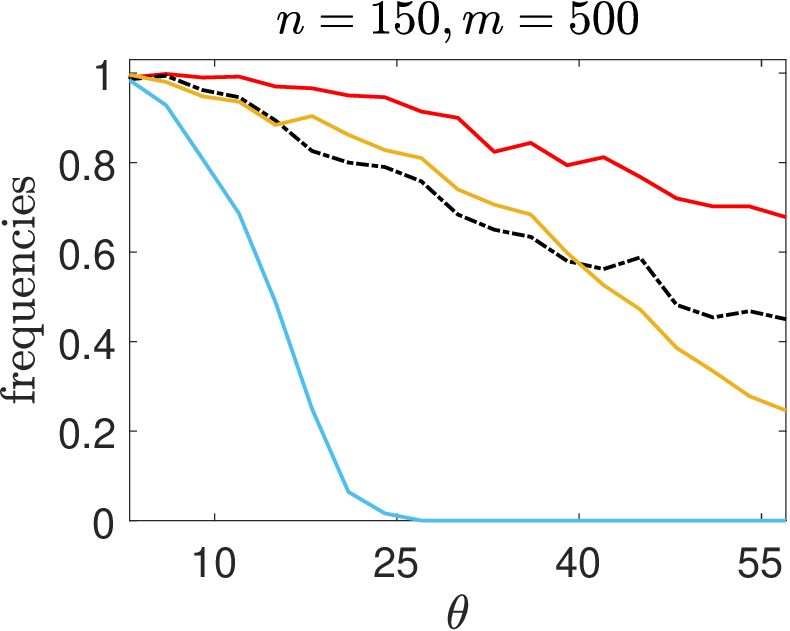}
        \vspace{0.2cm}
       \includegraphics[scale=0.365]{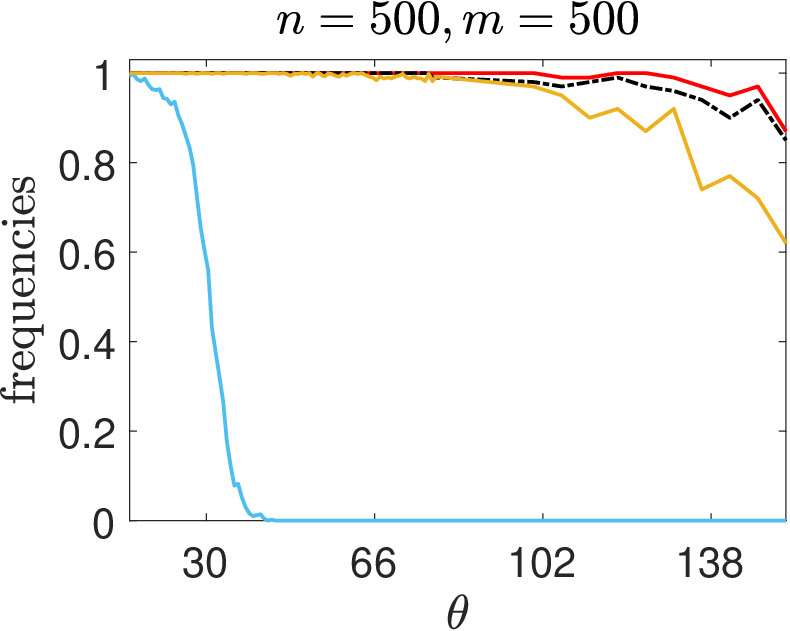}
    \end{minipage}%

\centering
\caption{\scriptsize{The relative frequencies of correctly selecting the number of factors for model \eqref{Models} under Scenario 2 with cross correlated errors (E3). In each panel, we use red curve for $FTCV_1$, black dash dotted curve for $FTCV_{10}$, blue curve for $IC_\p$, and yellow curve for $Ladle$. }}
\label{fig6}
\end{figure}

 In all subsequent simulation studies, we select $p$ from $\{1,2,..., p_{max}\}$ where $p_{max}$ is fixed at $8$ as in \cite{bai2002determining}. The combinations of $n$ and $m$ are systematically explored, with $n$ taking values from $\{40,150,500\}$ and $m$ from $\{40,150,500\}$. All suggested procedures for comparisons are replicated 500 times. Besides, to account for the effect of the signal-to-noise ratio on different methods, we allow the parameter $\theta$ to vary across intervals, dependent on diverse configurations of sample size $n$, dimensionality $m$, and idiosyncratic errors.
The frequencies of correct estimation of the number of factors over 500 replications at different noise levels are displayed in Figures \ref{fig1}-\ref{fig6}. Specifically, Figure \ref{fig1}-\ref{fig2} report the simulation results for Scenario 1 and Scenario 2 with independent Gaussian errors (E1). To evaluate the robustness of our proposed method, we extend our investigation to scenarios involving heteroskedastic and cross-correlated errors. Figure \ref{fig3}-\ref{fig4} present the results for Scenario 1 and Scenario 2 with heteroskedastic errors (E2), and Figure \ref{fig5}-\ref{fig6} report findings for both models with cross-correlated errors (E3).

\par In general, the simulation results consistently indicate improved performance for all methods as both $n$ and $m$ increase, providing empirical support for the asymptotic consistency of these techniques. All methods exhibit precision in estimating $p_0$ when $\theta$ is small, but their accuracy diminishes as the noise $\theta$ increases. Notably, $Ladle$ tends to outperform $IC_\p$, possibly attributed to its incorporation of both eigenvalues and eigenvectors in the estimation process. Remarkably, both $FTCV_1$ and $FTCV_{10}$ demonstrate superior performance across various scenarios, and this advantage becomes more pronounced as $m$ increases. They exhibit a notably higher frequency of correct selection, especially in cases where $\theta$ is large. This suggests their efficacy in accurate selection, even when the functional factors are uncorrelated and their contribution to the variation of the variables is low, which is particularly common in economic and financial data. Nevertheless, $Ladle$ dominates other selection methods when $n\gg m$ in some scenarios. Meanwhile, $FTCV_1$ and $FTCV_{10}$ continue to exhibit similar trends even when the idiosyncratic errors are heteroskedastic or cross-correlated. This consistency underscores the robustness of our method across diverse conditions.

\section{Application to Asset Returns} \label{sec6}

\subsection{Empirical Application to the Effect of Market Volatility on U.S. Treasury Yields}
In the literature, there is broad agreement that the U.S.  Treasury structure could be well explained by three factors, which are commonly interpreted as the level, slope and curvature factor \citep{cochrane2005bond,diebold2006forecasting}. To analyze how these three factors change with economic conditions, \cite{pelger2022state} built a state-varying factor model to illustrate the influences. Conditioned on different macro-economic state variables, the exposure to the three factors are actually captured by the factor loadings. In this section, we apply our semiparametric factor model to empirically study the influence of market volatility on the U.S. Treasury Yields and especially focus on the selection of the number of common factors.

The dataset contains daily data of the U.S. Treasury Securities Yields covering the period from October 16, 2018 to December 1, 2022. The terms range from 1, 2, 3, 6 months to 1, 2, 3, 5, 7, 10, 20, 30 years. Since the records for the 2-month treasury bill began at October 16, 2018, we set this date as the starting date to incorporate the 2-month treasury bill into our study. To examine the effect of market volatility on U.S Treasury Securities Yields, we use the Chicago Board Options Exchange Market Volatility Index (VIX) as the conditioning variable and transform the values by taking natural logarithms to account for the heavy tails. Meanwhile, the Treasury yields are centered in this study. After removing the dates with incomplete records, we end up with data for $m=12$ maturities and $n=1019$ observations. To investigate whether the results differ from different periods, we compute the estimated number of common functional factors with four approaches for various one-year and two-year periods. The maximum number of functional factors is set to be 8 as in the simulation studies.

\begin{figure}[htb]
\centering
        \includegraphics[scale=0.39]{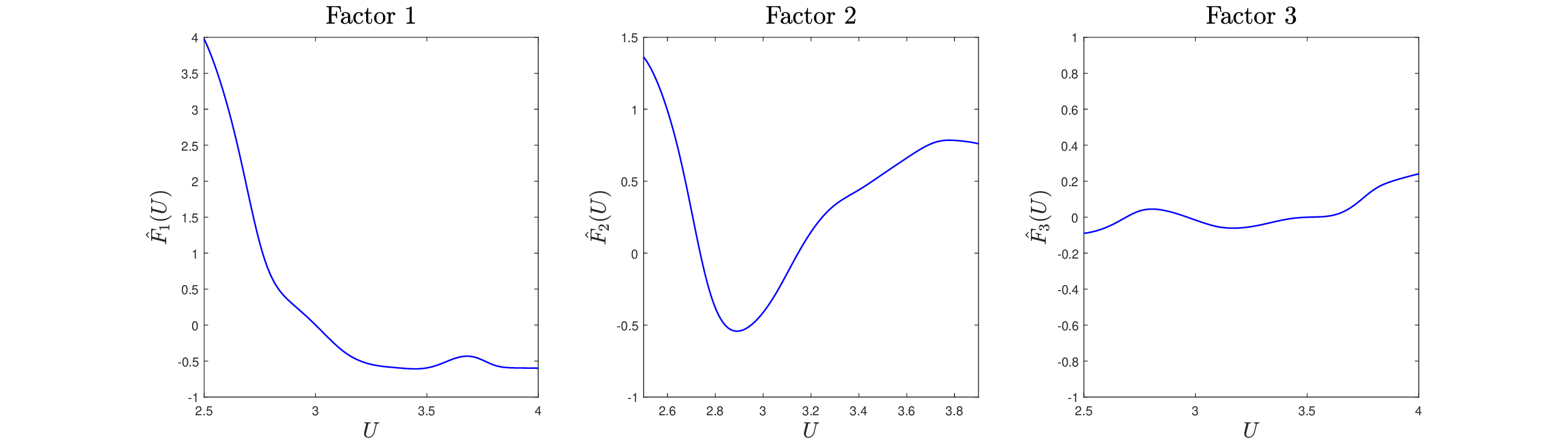}
        \vspace{-1cm}
\caption{\scriptsize{Empirical estimate of three common functional factors}}
\label{fighatF}
\end{figure}

Table \ref{table1} reports the estimated results for all the estimators. The results obtained by $IC_\p$ are suspicious and quite different from those obtained through other methods, which suggests there is only one or even no common functional factors affecting the changes of treasury yields.
$Ladle$ provides the closest results to those by FTCV, while it selects fewer factors and seems more unstable. Our FTCV analysis indicates the presence of three common functional factors across all periods considered, except for the period from October 2020 to October 2022 and those encompassing the first half of 2020. The occurance of four estimated common functional factors might be attributed to the impact of the COVID-19 outbreak. Moreover, to intuitively observe the changes in the common functional factors, we give the plots for the estimated common functional factors $\hat F_1(\cdot)$, $\hat F_2(\cdot)$ and $\hat F_3(\cdot)$. Representative plots for the three functional factors are displayed in Figure \ref{fighatF}. These findings are in accordance with the results presented  in the majority of literature, including \citep{pelger2022state}.

\begin{table}[ht!]
\centering
\caption{The estimated number of common functional factors} \label{table1}
\vspace{0.4cm}
\begin{small}
\begin{tabular}{cccccc}
\toprule
&  & $FTCV_1$ & $FTCV_{10}$ & $Ladle$ & $IC_{\p}$ \\
\cline{1-6}
\multirow{11}*{One-year Periods} &2018 Oct - 2019 Oct  & 3 & 3  &1 & 0\\
&2019 Jan - 2020 Jan &3 & 3 & 2 & 0\\
&2019 May - 2020 May  &  4 & 4 & 2 & 0\\
&2019 Oct - 2020 Oct  & 4  & 4 &3 & 0\\
&2020 Jan - 2021 Jan    & 4 & 4 &3 & 0\\
&2020 May - 2021 May   &  3 & 3  & 1 & 1\\
&2020 Oct - 2021 Oct   & 3 & 3 &2 & 0\\
&2021 Jan - 2022 Jan    & 3 & 3 &  0 & 0\\
&2021 May - 2022 May   &4  & 3  & 1 & 0\\
&2021 Oct - 2022 Oct   & 3 & 3 &  3 & 0\\
&2022 Jan - 2022 Dec   & 3 &3 &   2 & 0\\
\cline{1-6}
\multirow{8}*{Two-year Periods} &2018 Oct - 2020 Oct  & 4 & 4  &1 & 0\\
&2019 Jan - 2021 Jan & 4 & 4 & 1 & 0\\
&2019 May - 2021 May  & 4  & 4 & 3 & 0\\
&2019 Oct - 2021 Oct  & 4  &4  & 2 & 0\\
&2020 Jan - 2022 Jan   & 4 & 3  & 2& 0\\
&2020 May - 2022 May   & 3  & 3  & 2 & 0\\
&2020 Oct - 2022 Oct   & 4  & 4 & 3 & 0\\
&2021 Jan - 2022 Dec   & 3  & 3 &  2 & 0\\
\bottomrule
\end{tabular}
\end{small}
\end{table}

\subsection{Empirical Application to the Effect of Market Volatility on Stock Returns }
A number of previous studies have demonstrated that it is not uncommon to observe structural breaks or smooth changes (over time) in the second-order moment structure of economic or financial time series variables. In recent years, there has been increasing interest in studying how the covariance structure responds to the changes of some conditioning variables. Pairwise correlations, which serve as significant components of covariance matrices, play a crucial role in characterizing the co-movements and associations among variables. Indeed, the cross-correlations in financial markets are not constant over time. Existing literature has highlighted that the cross-correlations computed during volatile periods are significantly larger than those during calm periods \citep{longin1995correlation,chesnay2001does,ang2002asymmetric,jiang2016asymmetric}. It is noteworthy that \cite{jiang2016asymmetric} provided a strong evidence for such finding based on the semiparametric factor model, and employed an information criterion to select the number of common factors. In this study, by allowing the correlations to evolve smoothly with certain conditioning variable(s), we aim to study the performance of our FTCV method in identifying the number of common functional factors, by avoiding predetermination of penalty functions or tunning parameters.

We have gathered a dataset comprising the daily adjusted closing prices of the Dow30 components and the Chicago Board Options Exchange Market Volatility Index (VIX). The data covers the period from 4 January 2010 to 30 December 2022. After removing the companies with missing stock prices throughout the entire period, we have retained records for 28 companies. Similar to the previous example, we utilize the VIX as a measure of market volatility. However, our objective is to investigate the number of common factors that drive the correlations between the returns of these stocks. In this study, we estimate the pairwise conditional correlations between the stock returns based on the method introduced by \cite{jiang2016asymmetric}, resulting in a total of $m = \frac 1 2 \times 28\times(28-1) = 378$ correlations and $n=4278$ observations.  As usual, the returns of 28 stocks are computed as the first difference of the logarithm of the adjusted closing prices. Additionally, we employ devolatilized returns to eliminate the impact of leverage. This involves assuming that the raw returns follow an AR(1) + GARCH(1,1) process.

Table \ref{table2} displays the estimated number of common functional factors for the cross-correlation coefficients.
\begin{table}[ht!]
\centering
\caption{The estimated number of common functional factors for conditional correlations} \label{table2}
\vspace{0.4cm}
\begin{small}
\begin{tabular}{cccccc}
\toprule
Periods   & $FTCV_1$ & $FTCV_{10}$ & $Ladle$ & $IC_{\p}$ \\
\cline{1-5}
2006 Jan - 2022 Dec & 1 & 1  &1 & 1\\
2010 Jan - 2022 Dec & 1 & 1  &1 & 1\\
2014 Jan - 2022 Dec & 1 & 1  &1 & 1\\
2018 Jan - 2022 Dec & 1 & 1  &1 & 1\\
\bottomrule
\end{tabular}
\end{small}
\end{table}
The results unequivocally indicate that all four methods converge on the selection of a single common functional factor ($\hat p=1$), offering compelling evidence for the role of one function of market volatility as a driving force behind changes in the comovements of Dow30 returns. This conclusion aligns with the findings presented in a previous study by \cite{jiang2016asymmetric}.

\section{Concluding Remarks} \label{sec7}
Overall, our research contributes a novel approach for estimating the number of common functions in semiparametric factor models. These common functions play a crucial role in explaining the influence of exogenous variables on changes in response variables. In order to provide a robust and fully data driven strategy, we have proposed the Functional Twice Cross-Validation (FTCV) method for this purpose. This method ensures the orthogonality of the common factors and automatically selects the optimal number by balancing model complexity and stability. Through extensive simulation studies, we have demonstrated that our proposed method exhibits superior performance compared to existing methods in most cases, especially in datasets with large dimension $m$. Furthermore, we have also illustrated the practical application of our method by examing the number of common functional factors influencing the U.S. Treasury Yields and the cross correlations between Dow30 returns. By specifying the market volatility as the driving force, our findings align closely with the conclusions drawn in numerous existing studies.

\bibliographystyle{dcu}
\bibliography{NumOfFunFactor_ArXiv_version}



\end{document}